\documentclass[aps,prd,twocolumn,superscriptaddress,showpacs]{revtex4-1}

\usepackage[tight]{subfigure}
\usepackage{amsmath}
\usepackage{graphicx}
\usepackage{wasysym} 
\usepackage{color}
\usepackage{verbatim}

\newcommand{\gccm}{\ensuremath{\mathrm{g} \, \mathrm{cm}^{-3}}}

\newcommand{\kms}{\ensuremath{\mathrm{km\, s^{-1}}}}

\newcommand{\nuc}[2]{\ensuremath{\mathrm{^{#1}#2}}}
\newcommand{\ions}[2]{#1\,{\sc #2}}

\newcommand{\msun}{\ensuremath{\mathrm{M}_\odot}}
\def\MCh{$M_\mathrm{Ch}$}
\def\lesssim{\mathrel{\hbox{\rlap{\hbox{\lower4pt\hbox{$\sim$}}}\hbox{$<$}}}}

\def\gtrsim{\mathrel{\hbox{\rlap{\hbox{\lower4pt\hbox{$\sim$}}}\hbox{$>$}}}}

\begin{document}

\title{Towards an understanding of Type Ia supernovae\\
  from a synthesis of theory and observations}

\author{W.~Hillebrandt}
\affiliation
{ Max-Planck-Institut f\"ur Astrophysik,
  Karl-Schwarzschild-Str. 1, 
  D-85741 Garching, Germany
} 
\author{M.~Kromer}
\affiliation
{  Max-Planck-Institut f\"ur Astrophysik,
  Karl-Schwarzschild-Str. 1, 
  D-85741 Garching, Germany
} 

\author{F.~K.~R{\"o}pke}
\affiliation
{ Institut f{\"u}r Theoretische Physik und Astrophysik, 
  Universit{\"at} W{\"u}rzburg, 
  Am Hubland, 
  D-97074 W{\"u}rzburg, Germany
} 

\author{A.~J.~Ruiter}
\affiliation
{ Max-Planck-Institut f\"ur Astrophysik,
  Karl-Schwarzschild-Str. 1, 
  D-85741 Garching, Germany
} 

\begin{abstract}

Motivated by the fact that calibrated light curves of Type Ia
supernovae (SNe~Ia) have become a major tool to determine the
expansion history of the Universe, considerable attention has been
given to, both, observations and models of these events over the past
15 years. Here, we summarize new observational constraints, address
recent progress in modeling Type Ia supernovae by means of
three-dimensional hydrodynamic simulations, and discuss several of the
still open questions. It will be be shown that the new models have
considerable predictive power which allows us to study observable
properties such as light curves and spectra without adjustable
non-physical parameters. This is a necessary requisite to improve our
understanding of the explosion mechanism and to settle the question of
the applicability of SNe~Ia as distance indicators for cosmology. We
explore the capabilities of the models by comparing them with
observations and we show how such models can be applied to study the
origin of the diversity of SNe~Ia.
\end{abstract}

\pacs{ 97.60.Bw, 26.50.+x, 26.30.Ef, 26.30.-k, 97.60.-s, 97.80.-d}

\maketitle

\section{Introduction}
\label{sec:intro}

Today, Type Ia supernovae (SNe~Ia for short) play a somewhat ambiguous
role in astrophysics.
On the one hand, it is their relative homogeneity that
caused their use as distance indicators in observational cosmology. On
the other hand, this evoked an enormous interest resulting in a rather
comprehensive observational survey of SNe~Ia that over the last decade
clearly revealed sub-classes with diverging properties and variability
among these objects. While the notion of homogeneity inspired the
model of SNe~Ia being explosions of Chandrasekhar-mass white dwarfs (WDs), 
the newly discovered heterogeneity of the class suggests multiple 
progenitors and/or explosion mechanisms. 

After the pioneering work by
\citet{arnett1969a} numerical simulations have been instrumental in
modeling supernovae. Until the 1990s this approach was restricted to
one spatial dimension which prevented a realistic treatment of the
multi-dimensional burning mechanism in these objects. However,
parametrized models of that time, notably the W7 model of \citet{nomoto1984a},
still set a standard in the field and are widely used in the
interpretation of observational data. In the 1990s the first
multi-dimensional SN~Ia simulations emerged. Together with earlier work on
one-dimensional models, they are reviewed by
\citet{hillebrandt2000a}. Here, we report on developments in the last
decade, focusing on work associated with the supernova group at the
Max-Planck-Institute for Astrophysics, Garching, but putting it into
context with other work.

While the modeling of the explosion physics has made substantial
progress (in particular with the introduction of multi-dimensional
simulations), the question of the progenitor system of SNe~Ia remains
a fundamental problem. There is wide consensus that these events are
due to thermonuclear explosions of WDs \citep{hoyle1960a},
most likely composed of carbon and oxygen. This was recently confirmed
by \citet{nugent2011a} and \citet{bloom2012a} who on the basis of
early time observations concluded that the exploding object in SN~Ia
2011fe was a compact star. The question of how it reaches an explosive
state, however, is more complicated. As single WDs are
unconditionally stable, some kind of interaction with another star is
necessary to explain the supernova. Unfortunately, attempts to
identify this star beyond doubt have failed so far. Collisions with
compact objects in globular clusters \citep{rosswog2008a,
rosswog2009a, raskin2009a} lead to atypical events or fall short of
explaining the SN~Ia rate. Thus, although such events may occur in
Nature, the bulk of SNe~Ia is more likely to be associated with
stellar binaries. The nature of the binary companion, however, is
still unclear. Traditionally, two classes of potential binary
progenitor systems have been distinguished -- the single-degenerate
progenitor channel, in which the companion is a normal star, and the
double-degenerate channel with two WDs interacting and merging. At
present it is unclear whether one of these possibilities is
exclusively realized in Nature or whether both contribute to the class
of SNe~Ia.

This problem has been approached from different
perspectives. Observational data becomes increasingly constraining for
the physical mechanism of SNe~Ia and a brief overview of the current
status is given in Sect.~\ref{sec:observations}. In addition, the rate
at which SNe~Ia occur and the distribution of delay times between
formation of the progenitor systems and supernova explosions can help
to identify the dominant progenitor channel(s).  These data can be
compared with predictions from binary population synthesis
calculations. We discuss this approach and recent results in detail in
Sect.~\ref{sec:progenitors}. Another possibility is to follow
different explosion scenarios in hydrodynamic simulations. Combined
with radiative transfer calculations these predict observables that
can be directly compared to SN~Ia observations. Over the last decade
substantial progress in this approach was possible due to fully
multi-dimensional treatment that allows to reduce the free parameters
involved in describing the physics and thus improve the predictivity
of the models. We discuss recent results in Sect.~\ref{sec:models},
where the modeling approaches and the implementation in numerical
simulations are briefly outlined followed by the presentation of models
that potentially can account for normal SNe~Ia
(Sect.~\ref{sec:normals}), while other models seem to reproduce
peculiar subclasses (Sect.~\ref{sec:peculiars}). We emphasize that
this is a way of presenting our models and discussing the
results. However, the assignment of models to different subclasses is
not necessarily unique, but it is chosen here to point out the
possibility to model a wide variety of SNe~Ia when considering
different progenitor scenarios and explosion mechanisms.

\section{Observations}
\label{sec:observations}

The efforts to systematically obtain observational data of SNe~Ia have
gained tremendous momentum during the past 15 years. This is primarily
a result of their unequaled potential to act as `standardizable' 
candles for the measurement of the cosmological expansion rate and its 
variation with look-back time
\citep{phillips1993a,riess1996a,branch1998a,schmidt1998a,riess1998a,
perlmutter1999a,tonry2003a,riess2004a,astier2006a,riess2007a,mignone2008a,
benitez2012a} (see also \citet{goobar2011a} for a recent
review). The discovery that the Universe entered into a phase of
accelerated expansion at a redshift of around 0.5, due to the action
of some unknown form of `dark energy', was awarded with the Nobel
Prize in Physics in 2011 to Saul Perlmutter, Adam Riess, and Brian Schmidt.  

For theorists, this development presents both a challenge, to help 
to understand the correlations among the observables, and an opportunity, 
to use the wealth of new data to constrain the zoo of existing 
explosion models. There exist a number of excellent reviews about SNe~Ia
observations in general \citep{filippenko1997b,leibundgut2000a,
leibundgut2008a}, their spectral properties \citep{filippenko1997a,
branch2009a}, and photometry in the IR and optical bands 
\citep{meikle1996a,meikle2000a,leibundgut2003a}. Here, we highlight 
those aspects of SN~Ia observations that most directly influence 
theoretical model building at the current time.
 
\subsection{General properties}
\label{sec:properties}

The classification of SNe~Ia is based on spectroscopic features: the
absence of hydrogen absorption lines, distinguishing them from Type II
supernovae, and the presence of strong silicon lines in their early and
maximum-light spectra, classifying them as Type Ia's 
\citep{wheeler1990a}.

The spectral properties, absolute magnitudes, and light-curve shapes
of the majority of SNe~Ia are remarkably similar, exhibiting only
small spectroscopic and photometric differences \citep{branch1998a}. 
It was believed until recently that approximately 85\% of all observed 
events belong to this class of `normal' \citep{branch1993a} SNe~Ia, 
represented for example by SN 1972E, SN 1994D, or SN 2005cf. However, 
recent studies show that the peculiarity rate can be as high as 30\% 
as suggested for instance by \citet{li2011a}.

The optical spectra of normal SNe~Ia contain neutral and
singly-ionized lines of Si, Ca, Mg, S, and O at maximum light,
indicating that the outer layers of the ejecta are mainly composed of
intermediate mass elements \citep{filippenko1997a}. Permitted
\ions{Fe}{ii} lines dominate the spectra roughly two weeks after
maximum when the photosphere begins to penetrate Fe-rich ejecta
\citep{harkness1991a,branch2009a}. In the nebular phase, beginning
approximately one month after peak brightness, forbidden
\ions{Fe}{ii}, \ions{Fe}{iii}, and \ions{Co}{iii} emission lines
become the dominant spectral features.  Some \ions{Ca}{ii} remains
observable in absorption even at late times
\citep{filippenko1997a}. The decrease of Co lines and the relative
intensity of \ions{Co}{iii} and \ions{Fe}{iii} give evidence that the
light curve tail is powered by radioactive decay of $^{56}$Co
\citep{kuchner1994a}(see also \citet{truran1967a,colgate1969a}).

The early spectra can be explained by resonant scattering of a thermal
continuum with P Cygni-profiles whose absorption component is
blue-shifted according to ejecta velocities of up to about
25,000\,\kms, rapidly decreasing with time.
Different lines have different expansion velocities
\citep{patat1996a,stehle2005a,garavini2007a}, suggesting a layered 
structure of the explosion products.

Photometrically, SNe~Ia rise to maximum light in a period of
approximately 18 to 20 days \citep{riess1999a,strovink2007a,
conley2006a,hayden2010a} reaching
\begin{equation}
M_{\rm B} \approx M_{\rm V} \approx -19.30 \pm 0.03 + 5 \log(H_0/60)
\end{equation}
with a dispersion of $\sigma_M \le 0.3$ \citep{hamuy1996a}. It is
followed by a first rapid decline of about three magnitudes in a
matter of one month. Later, the light curve tail falls off in an
exponential manner at a rate of approximately one magnitude per
month. In the $I$- and near-infrared bands, normal SNe~Ia rise to a second 
maximum approximately 20 days after the first one \citep{meikle2000a}.
Typical $^{56}$Ni masses inferred from their bolometric light curves
are in the range from 0.3 to 0.9\,\msun\ for normal SNe~Ia
(e.g., \citet{stritzinger2006a}).

It is especially interesting that the two most abundant elements in
the universe, hydrogen and helium, so far have not been unambiguously
detected in the spectra of normal SNe~Ia
\citep{filippenko1997a,leonard2007a} (but see \citet{meikle1996a} and
\citet{mazzali1998b} for a possible identification of He, and 
\citet{hamuy2003a} and \citet{dilday2012a} for an identification 
of H in individual interacting peculiar objects). 
Also, there are no indications yet 
for radio emission \citep{chomiuk2011a}, including the rather 
nearby supernova SN 2011fe \citep{chomiuk2012a,horesh2012a}.  

\subsection{Diversity and correlations}
\label{sec:observed_diversity}

Early suggestions \citep{pskovskii1977a,branch1981a} that the existing
inhomogeneities among SN~Ia observables are strongly intercorrelated
are now established beyond doubt \citep{hamuy1996a,filippenko1997a}. 
\citet{branch1998a} summarizes the correlations between spectroscopic 
line strengths, ejecta velocities, colors, peak absolute magnitudes, 
and light curve shapes that were known at that time. Roughly speaking, 
SNe~Ia appear to be arrangeable in a one-parameter sequence according 
to explosion strength, wherein the weaker explosions are less
luminous, redder, and have a faster declining light curve and slower 
ejecta velocities than the more energetic events. Based on these
findings  \citet{mazzali2007a} argue that a single explosion scenario,
possibly a delayed detonation (see Sect.~\ref{sec:normals_from_MCh}),
may explain most SNe~Ia. However, more recent (and better) data
challenge this conclusion, as will be discussed below.

\begin{figure}
  \centering
  \includegraphics[width=\linewidth]{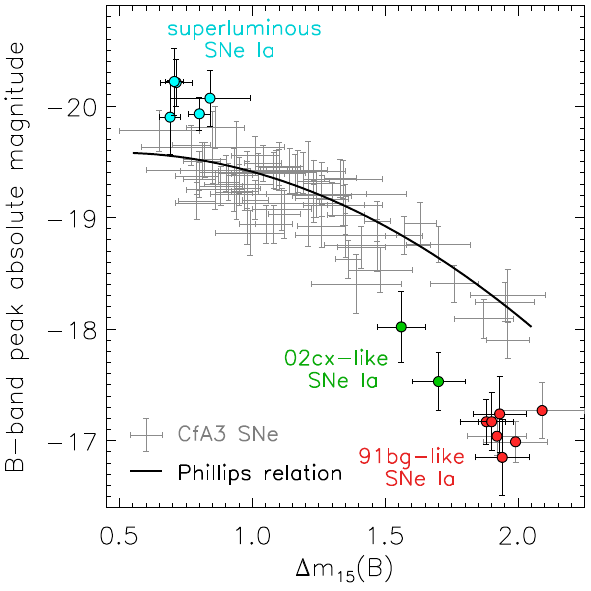}
  \caption{Observational diversity of SNe~Ia in $B$-band decline
    rate $\Delta m_{15}(B)$ and $B$-band peak absolute magnitude.
    Normal SNe~Ia (shown in grey, data taken from \citet{hicken2009b})
    follow the Phillips relation \citep{phillips1999a}. 1991bg-like
    SNe (shown in red, data taken from \citet{taubenberger2008a}),
    and 2002cx-like SNe (shown in green, data taken from 
    \citet{phillips2007a}) are subluminous with respect to the
    Phillips relation. Superluminous SNe~Ia (shown in cyan, data from
    \citet{taubenberger2011a}) are almost one magnitude brighter in 
    $B$-band than normal SNe~Ia with a comparable $B$-band decline rate.}
  \label{fig:snia_diversity}
\end{figure}

The relation between the width of the light curve around maximum and
the peak brightness (brighter supernovae decline more slowly) is the
most prominent of all correlations (Fig.~\ref{fig:snia_diversity};
\citet{pskovskii1977a,phillips1993a}). 
Parameterized either by the decline rate $\Delta m_{15}$ 
\citep{phillips1993a,hamuy1996a}, a `stretch parameter' 
\citep{perlmutter1997a}, or a multi-parameter nonlinear fit in 
multiple colors \citep{riess1996a}, it was used to renormalize the 
peak magnitudes of a variety of observed events, substantially 
reducing the dispersion of absolute brightnesses (see,
e.g., \citet{leibundgut2008a} and \citet{goobar2011a} for recent
reviews). This correction procedure is a central ingredient of all
current cosmological surveys that use SNe~Ia as distance indicators
\citep{kowalski2008a,amanullah2010a,lampeitl2010a,guy2010a,conley2011a,
suzuki2012a}.

However, there are supernovae, classified as Type Ia, which violate
this correlation. SN 1991bg, SN 1992K, SN 1999by and SN 2005bl are
well-studied examples for red, fast, and subluminous supernovae with a
typical $\Delta m_{15}$ value of about 1.8 and $B$-band peak
absolute magnitudes around -17, roughly one magnitude fainter than
their `normal' counterparts 
\citep{filippenko1992a,leibundgut1993a,hamuy1994a,turatto1996a,
  garnavich2004a,taubenberger2008a}. 
Their $V$, $I$, and $R$-band light curves decline unusually quickly,
skipping the second maximum in $I$, and their spectra show a high
abundance of intermediate mass elements (including \ions{Ti}{ii}) with
low expansion velocities but only little iron.  Models for the nebular
spectra and light curve of SN 1991bg consistently imply that the total
mass of $^{56}$Ni in the ejecta was very low ($\sim
  0.07$\,\msun \citep{mazzali1997a}), a typical value for this class being
$\sim 0.1$\,\msun. In addition, there is also evidence for unburnt C
and O in their early spectra, in contrast to normal SNe~Ia. These
`subluminous' explosions make up for about 15\% (or more) of all
SNe~Ia \citep{li2011a}.

The prototype of a second group of subluminous SNe~Ia is SN~2002cx
\citep{li2003a,jha2006a}. Here, again, the mass of $^{56}$Ni, as 
estimated from `Arnett's rule' \citep{arnett1982a}, is low, around 
0.2 \,\msun\ only. The spectra show narrow lines, indicating low
ejecta velocity and low kinetic energy. Other supernovae belonging
to this class include SN 2005hk \citep{phillips2007a,sahu2008a}, 
SN2008ge \citep{foley2010b}, and SN 2009ku \citep{narayan2011a}.
According to \citet{li2011a} they contribute about 5\% of all SNe~Ia.
Even 300 days after the explosion, the ejecta of members of 
this group are not transparent, but show emission from a narrow 
region in velocity space (less than 1000\,\kms\, \citep{jha2006a}).

Finally, transients even fainter than 1991bg-like SNe have been
observed, SN~2005E \citep{perets2010a} or SN~2005cz
\citep{kawabata2010a} being examples. They are Ca-rich fast decliners,
their spectra resemble more SNe Ib than SNe~Ia, i.e., they show He but
little O and Si in their early-time spectra, and their decline rates
are similar to those of SNe~Ic. They are found in old stellar
populations, however, and the discussion is open whether they are
thermonuclear explosions or core-collapse supernovae
\citep{waldman2011a,maeda2012a,kasliwal2012a}.

At the other end of the luminosity function, SN 1991T is often
mentioned as a striking representative of bright, energetic events
with broad light curves
\citep{phillips1992a,jeffery1992a,filippenko1992a,
  ruiz-lapuente1992a,spyromilio1992a}. Rather than the expected
\ions{Si}{ii} and \ions{Ca}{ii}, its early spectrum displayed
high-excitation lines of \ions{Fe}{iii} but returned to normal a few
months after maximum light. But recently other SNe~Ia were found which
are even more luminous than SN 1991T, with decline rates that put them
well above the Phillips relation by almost one magnitude in the
$B$-band, prototypical examples being SN 2006gz and SN~2009dc
\citep{howell2006a,hicken2007a,
  yamanaka2009a,tanaka2010a,silverman2011a,taubenberger2011a}.
By now, seven objects that may belong to this subclass have been
discovered and they may contribute up to about 9\% of all SNe~Ia
\citep{li2011a}.  In addition to their high luminosity, 2 to 3 times
higher than normal SNe~Ia, they are characterized by a slow decline 
($\Delta m_{15}(B) \sim 0.8$), a long rise time ($\geq 23$ days), 
low ejecta velocities, and prominent \ions{C}{ii} absorption
features, while other properties of their early-time spectra are 
similar to those of normal SNe~Ia. If the luminosity at peak would 
come exclusively from the decay of $^{56}$Ni
the Ni-mass of SN~2009dc would be around 1.5 to
1.8\,\msun\ \citep{taubenberger2011a,kamiya2012a}, exceeding the
Chandrasekhar mass (see, however, \citet{hachinger2012a} for an
alternative scenario).  The various sub-classes are illustrated in
Fig.~\ref{fig:snia_diversity}.

From early on, peculiar events like SN 1991T and SN 1991bg were 
suggested to belong to different subgroups of SNe~Ia than the normal 
majority, created by different explosion mechanisms
\citep{mazzali1997a,filippenko1992a,fisher1999a} although the overall 
SN~Ia luminosity function seems to be rather smooth, with a shallow
increase from an absolute $R$-band magnitude of -17 to -19, followed 
by a steep decline to -19.5 \citep{li2011a} (thus leaving out 
09dc-like events), indicating that `normal' SNe~Ia are essentially 
the brightest, with Ni masses around 0.6\,\msun\ while the full 
class may contain a large number of undetected subluminous SNe~Ia. 

\subsection{SNe~Ia and their host galaxies}
\label{sec:hosts}

There is mounting evidence that SN~Ia observables are correlated
with their host stellar population \citep{branch1998a} and there
are recent investigations demonstrating the dependence of
supernova properties on global characteristics of their hosts
\citep{sullivan2006a,howell2009a,hicken2009a,kelly2010a,sullivan2010a}. 
For instance, SNe~Ia in red or early-type galaxies show, on average,
slower ejecta velocities, faster light curves, and are dimmer by 
$\approx 0.2$ to 0.3 mag than those in blue or late-type star-forming 
galaxies \citep{hamuy1996a,branch1996a,gallagher2005a,sullivan2006a,
lampeitl2010a}. Moreover, SNe~Ia seem to have lower ejecta 
velocity in high-mass host galaxies \citep{foley2012a}. On the
other hand side, SNe Ia at low and high redshift seem to have 
similar spectral evolution
\citep{blondin2006a,bronder2008a}.

The SN~Ia rate per unit stellar mass is nearly a factor of 20 higher in 
late-type galaxies than in early-type ones and depends inversely on
the host galaxy's mass \citep{mannucci2005a,li2011b,graur2012a}. 
The rate seems to be lower in galaxy bulges than in spiral arms.
These findings indicate that there might be a population of
progenitors with large delay time 
\citep{cappellaro1999a,perrett2012a}. Also, the outer regions of 
spirals appear to give rise to similarly dim SNe~Ia as ellipticals 
whereas the inner regions harbor a wider variety of explosion
strengths \citep{wang1997a}. 

\subsection{Summary: observational requirements for explosion models}
\label{sec:requ}

To summarize the main observational constraints, any viable scenario
for the SN~Ia explosion mechanism has to satisfy the following
(necessary but probably not sufficient) requirements:
\begin{enumerate}
\item Agreement of the ejecta composition and velocity with observed
spectra and light curves. In general, the explosion must be
sufficiently powerful (i.e., produce enough $^{56}$Ni) and produce a
substantial amount of high-velocity intermediate mass elements in the
outer layers. Furthermore, the isotopic abundances of ``normal'' SNe~Ia 
must not deviate significantly from those found in the solar system.
\item Robustness of the explosion mechanism. In order to account for
the homogeneity of normal SNe~Ia, the standard model should not give
rise to widely different outcomes depending on the fine-tuning of
model parameters or initial conditions.
\item Intrinsic variability. While the basic model should be robust
with respect to small fluctuations, it must contain at least one
parameter that can plausibly account for the observed sequence of
explosion strengths. However, this could in principle also be achieved 
by allowing for different progenitor channels.
\item Correlation with progenitor system. The explosion strength
parameter must be causally connected with the state of the progenitor
WD in order to explain the observed variations as a function
of the host stellar population. Moreover, there must be a sufficient 
number of progenitor systems such that the rate and delay-time
constraints are matched.
\end{enumerate}

\section{Progenitors}
\label{sec:progenitors}

While it is widely accepted that SNe Ia originate from explosions of
WDs that approach critical conditions such that burning can proceed
explosively (cf.~Sect.~\ref{sec:intro}), the manner in which these
conditions are achieved remains uncertain.  Almost certainly the WD
gains matter from a nearby stellar companion.  Until very recently,
the standard paradigm was the following: SNe Ia originate from
probably one, possibly two different formation channels which enable
the WD to reach critical conditions necessary for a thermonuclear
explosion to occur (see following subsections).

However, recent observations of SNe Ia have brought to light the
(previously shrouded) highly diverse nature of these objects (see
Sect.~\ref{sec:observations}).  When one considers all of the
necessary criteria that a progenitor model must satisfy in order to be
seen as a viable progenitor candidate (robust explosion mechanism,
ejecta stratification, velocities and nucleosynthesis, characteristic
peak luminosity and light-curve shape, absolute birth rates and delay
times), it is evident that reconciliation of the entire range of
observed characteristics of SNe Ia with a single progenitor scenario
is improbable.  If more than one progenitor scenarios are contributing
to the observed population of SNe Ia -- which is currently the
favoured view \citep{wang2012b} -- it is still unclear as to which
progenitor scenario(s) dominate(s).

Binary population synthesis models have been used for a few decades
now to estimate relative (and absolute) birthrates of various binary
formation channels that can lead to SNe Ia (see \citep{iben1984a} for
one of the most well-known early studies).  A powerful feature of
population synthesis models is that one is able to easily compute the
delay times of SNe Ia, which puts strict limits on the system age,
thus ruling out certain theoretical progenitor scenarios.  In
addition, the models enable one to reconstruct the entire evolutionary
history for all binaries of interest, which is critical for uncovering
evolutionary phases (e.g. mass transfer episodes) that might give rise
to observational features which could kill or confirm a given model
(see e.g.~\citet{ruiter2012a} in which the {\sc StarTrack} 
\citep{belczynski2008a} binary evolution synthesis code is used).

In terms of SN Ia rates and delay times, the results from different
population synthesis codes are found to vary quite a lot in some cases
and agree fairly well in others (\citet{nelemans2012a} and references
therein).  This is primarily due to the rather uncertain nature of
mass transfer and accretion in close binary stars which leads to
differing assumptions for the input physics in the various codes.  In
particular, for progenitors which undergo (quasi)stable mass exchange,
one must decide how the (donor) mass transfer rate is approximated in
the given binary and correspondingly, how the (accretor) retention
efficiencies are computed (see, e.g., \citep{han2004a,mennekens2010a}). Further
still, the manner in which matter is lost from the binary (carrying
away with it angular momentum) will also have an effect on the orbital
behaviour and subsequent binary evolution \citep{portegies2001a}.

The progenitor problem is still unsolved, though as previously
mentioned, it seems likely that at least two progenitor scenarios (and
possibly more explosion mechanisms) are required in order to explain
the observed SN Ia rate and delay time distribution \citep{maoz2012b}.
In the following sub-sections we review the most promising progenitor
scenarios (e.g. formation channels) which are thought to lead to SNe
Ia.

\subsection{Single-degenerate (Chandrasekhar-mass white dwarf)
  scenario} 
\label{sec:SD}

\subsubsection{Hydrogen-burning donors}

Often called the single degenerate (SD) scenario \citep{whelan1973a},
SD systems are detected in Nature (e.g.~RS Oph, \citep{hachisu2007a})
and were, in the past, widely thought to be the most promising SN Ia
progenitors.  In this scenario, the companion star is a main sequence
or giant-like star (possibly a helium-burning star; see
Sect.~\ref{sec:He-SDS}) that is overfilling its Roche-lobe,
transferring matter through the inner Lagrange point in a stable
manner to the companion carbon--oxygen (CO) WD.  If the mass transfer
proceeds within a certain range of rates (for example
\citep{nomoto2007a}), the donor material is accreted in a stable
fashion leading to efficient hydrogen-burning (mass accumulation) on
the WD, thereby increasing its (central) density.  When the density in
the center of a CO WD becomes high enough the carbon in the WD starts
to burn (see Sect.~\ref{sec:models}) which eventually leads to a
thermonuclear explosion, obliterating the WD and possibly imparting a
significant kick on the companion star (Sect.~\ref{sec:intro}).  This
critical density when carbon-burning can start is usually attained
when the WD approaches a critical mass -- the Chandrasekhar limit.

A typical formation pathway leading to the SD scenario involves 
an episode of unstable mass transfer followed by an episode of stable mass
transfer at a later stage.  A more specific example
(e.g. \citep{ruiter2009a}) is as follows: the initially more massive 
star (the primary) first loses its hydrogen-rich envelope on the asymptotic 
giant branch (AGB) when it fills its Roche-lobe and mass transfer is 
dynamically unstable.  This results in a common envelope, which
serves to bring the two stars to a smaller orbital separation
\citep{paczynski1976a}.  The post-common envelope binary 
comprises a (newly formed) CO WD and a (likely still on the main 
sequence) companion.  At some later stage, the companion then fills
its Roche-lobe (either while on the main sequence or as an evolved 
star), only this time mass transfer is stable, and the CO WD grows in
mass until it approaches the Chandrasekhar limit.  

\subsubsection{Helium-burning donors} 
\label{sec:He-SDS}

It is also possible that the WD may reach the Chandrasekhar mass, by
accreting from a helium-burning star donor rather than a hydrogen-rich
donor (e.g. \citep{solheim2005a}).  Such formation channels are
expected to be rare, and these progenitors have shorter evolutionary
timescales than the `canonical' SD scenario due to the larger zero-age
main sequence (ZAMS) mass of the secondary
\citep{iben1987a,ruiter2009a,wang2009c}.  Since they might also
harbour distinctly different physical (observable) properties, they
have been considered their own class by some authors (e.g.,
helium-rich (HeR) scenario, \citep{ruiter2011a}).

\subsection{Double-detonation (sub-Chandrasekhar-mass white dwarf)
  scenario}
\label{sec:prog-doub-det}

Another progenitor channel which has recently re-gained popularity
among the community is the sub-Chandrasekhar-mass scenario in which a
sub-Chandrasekhar-mass CO WD accretes stably from a companion and
never reaches the Chandrasekhar limit before exploding
\citep{iben1987a}.  Depending on the assumed mass transfer/accretion
rates and the mass of the CO WD, the WD is thought to be able to
accumulate (rather than burn) a layer of helium which may detonate
under the right physical conditions \citep{taam1980a,shen2010a}.
This shell-detonation -- if realized -- likely triggers a second
detonation in the sub-Chandrasekhar-mass WD, leading to a SN Ia
(`double-detonation' scenario, see Sect.~\ref{sec:sub_MCh}).

Such a scenario was investigated from a population synthesis
standpoint by \citet{tutukov1996a} in context of helium-rich donors,
and by \citet{yungelson1995a} in context of of symbiotic systems.
However, formation channels leading to the double-detonation scenario
via accretion from a hydrogen-rich companion are challenged by the
ability of the WD to efficiently accrete (and stably burn) hydrogen
and helium \citep{iben1996a,yungelson2000a}.  Nonetheless, a
double-detonation progenitor scenario might be readily realized in
Nature from sub-Chandrasekhar-mass WDs accreting from helium-burning
stars or helium-rich degenerate (or semi-degenerate) stars
\citep{iben1991a,tutukov1996a}.

In considering all potential helium-rich donors transferring mass to
sub-Chandrasekhar-mass CO WDs, \citet{ruiter2011a} investigated the
double-detonation scenario.  In that work, two characteristic
evolutionary channels were found (note that the authors only
considered SNe Ia to arise from systems where the total WD mass was
$\gtrsim 0.9$ \msun).  The typical formation channels were as
follows:\\
{\em Helium-burning star donors}.  This formation channel involves two
cases of unstable mass transfer (common envelopes) followed by a later
stage of stable mass transfer from the secondary to the CO WD.  The
initial primary star fills its Roche-lobe (unstably) while on the AGB
when the companion is still on the main sequence, resulting in a CO
WD--main sequence binary.  A second common envelope ensues when the
secondary star -- now evolved -- fills its Roche-lobe.  The
post-common envelope binary consists of a CO WD (left over from the
first common envelope) and a stripped core of a giant; a compact naked
helium-burning star.  Since the stars are already on a fairly close
orbit (due to two common envelopes), it does not take long for the
stars to be brought into contact.  The naked helium-burning star then
fills its Roche-lobe and transfers matter stably to the primary WD,
until the onset of the double-detonation. \\
{\em (Semi-)degenerate star donors}. 
A number of evolutionary pathways can lead to the formation of such a
progenitor, but the most common path also involves two common envelopes
(primary on the AGB, then secondary on the giant branch), followed by
a phase of stable mass transfer.  However, in this case, the
secondary's ZAMS mass is smaller than that of helium-burning star
donor case.  Thus, when the secondary loses its H-rich envelope in the
second common envelope, a degenerate (non-burning) naked helium core 
of a giant is left behind.  Once contact is achieved, the `helium-rich 
WD' transfers matter stably to the CO WD until the onset of the 
double-detonation.  

\subsection{Double degenerate mergers} 
\label{sec:DD}

Another scenario which might readily lead to SNe Ia is the merger of
two CO WDs where the total mass exceeds the Chandrasekhar limit
(`double degenerate' (DD) scenario, \citep{webbink1984a}). Along
with the SD scenario, the DD scenario has been a leading progenitor
candidate model.  The reason is owed partially (but not only) to the theoretical birth
rate calculations -- for which it historically does the best of any
progenitor scenario.  A number of population synthesis studies 
over the last few decades (e.g. \citep{iben1984a}) have demonstrated 
that mergers of CO WDs with a total
mass exceeding the Chandrasekhar limit might be frequent enough to
account for Galactic SN Ia rates, depending on e.g.\ the adopted
prescriptions for common envelope evolution  
\citep{yungelson1994a,ruiter2009a}.  Still, when considering 
{\em cosmological} SN~Ia rates as a function of delay time, the DD
model scenario, like other scenarios, often falls short of the 
observationally-recovered rates by at least a factor of a few 
\citep{maoz2010a,ruiter2011a,toonen2012a}. However, some of the most 
recently-measured delay time distributions indicate that DD merger 
rates might indeed be frequent enough to account for the bulk of
SNe~Ia at least in some stellar populations 
(\citep{graur2012a,ruiter2012a}, see Sect.~\ref{subsec:dtd_rates}).

There are a number of progenitor pathways that can lead to a CO-CO WD
binary.  A typical one would be the following \citep{ruiter2009a}: The
primary star fills its Roche-lobe when it is slightly evolved, and
mass transfer is stable to the companion (a second stable phase of
Roche-lobe overflow may follow a bit later, when the primary is an
evolved helium star).  The primary star then evolves into a CO WD.
The secondary star fills its Roche-lobe when it is an evolved star but
mass transfer is unstable to the WD, and a common envelope ensues.
The post-common envelope binary consists of a CO WD and a naked
helium-burning star.  Following this, a final phase of stable mass
transfer may occur whereby the slightly evolved helium star
(secondary) transfers matter to the primary WD.  Such a final phase of
mass transfer was found to be important in explaining the
peak-brightness distribution of SNe Ia (see Sect.~\ref{sec:assess}).
Once the WDs reach contact and the larger (less massive) one fills its
Roche-lobe, mass transfer must be unstable for a merger to occur.
This is the likely outcome for double CO WDs given their typical mass
ratios (see \citep{ruiter2010c}; see also \citet{toonen2012a} for 
specific examples of DD formation channels).

Despite favourable theoretical rate predictions, the DD scenario has
received a lot of criticism over the years.  Earlier calculations
predicted that the likelihood of thermonuclear explosion in a double
CO WD merger is rather unlikely.  Such a merger ($M_{\rm tot} \gtrsim$
\MCh) was thought to lead to disruption of the secondary WD which is
then accreted onto the primary.  The accretion would not lead to
central burning but rather burning in the outer layers of the WD,
where densities are lower, and the accreting CO WD would transform
into an oxygen-neon-magnesium WD \citep{saio1985a}.  For WDs of such
composition, electron captures become important at high central
densities, and as the WD approaches the Chandrasekhar mass it
collapses to form a neutron star in an accretion-induced collapse
(AIC, \citep{miyaji1980a}). Even if DD mergers do lead to SNe Ia, it
has been reiterated by some groups that, assuming the estimated
birthrates from population synthesis calculations are correct within a
factor of a few, all CO WD mergers (even those with $M_{\rm tot} <
1.4$ \msun) must lead to SNe Ia in order to match the observed rates
\citep{vankerkwijk2010a,badenes2012a}.  It is unclear whether
lower-mass mergers lead to thermonuclear explosions, let alone if such
explosions would produce enough \nuc{56}Ni.

Depending on the configuration of the binary system -- in particular
the mass ratio -- the merger may be somewhat quiescent as described
above, or it may be violent enough such that a prompt detonation in
the primary WD will occur.  These `violent mergers' are robustly found
to lead to a thermonuclear explosion, and they are described in
Sect.~\ref{sec:violent_mergers}.

\subsection{Other possible scenarios}

Other possible formation channels leading to SNe Ia have been
postulated in the literature.  For example: a potential scenario
involves the merger of a CO WD and the core of an AGB star during a
common envelope event \citep{sparks1974a,livio2003a}.  Such events are
expected to readily occur, however it is unclear whether such
`core-degenerate' mergers would lead to an immediate (or
delayed,\citep{ilkov2012a}) explosion that exhibits observational
signatures which match those of SNe Ia.

\subsection{Constraining progenitor models: delay times and rates}
\label{subsec:dtd_rates}

The delay time distribution (DTD) is the distribution of times in
which SNe Ia explode following a (hypothetical) burst of star
formation.  Knowing the DTD gives the age of the progenitor, which
places strong constraints on the different proposed progenitor
scenarios.  If the SN Ia rate is known in addition, then it becomes
possible to rule out theoretical formation channels.

Calculation of observationally-recovered DTDs involves many
assumptions, the most important being the assumed star formation
history of the supernova's host galaxy or local stellar population,
for which several techniques have been employed
\citep{foerster2006a,graur2012a}.  There are two emerging facts in the
literature: {\em i)} there is a population of `prompt' SNe Ia which
have delay times ${\lesssim} 500$ Myr and seem to comprise a
significant fraction of all SNe Ia, and {\em ii)} there are SNe Ia
which are `delayed', the seemingly-continuous DTD spanning up to a
Hubble time, with a characteristic (cosmic) DTD beyond $> 400$ Myr
that follows a power-law shape $t^{-1.2}$ \citep{maoz2010a}.  Such a
power-law (${\sim}t^{-1}$) is expected if the dominating timescale
leading to SNe Ia is set by gravitational radiation, as is the case
for DD mergers \citep{totani2008a}.

\begin{figure}
  \centering
  \includegraphics[width=9cm]{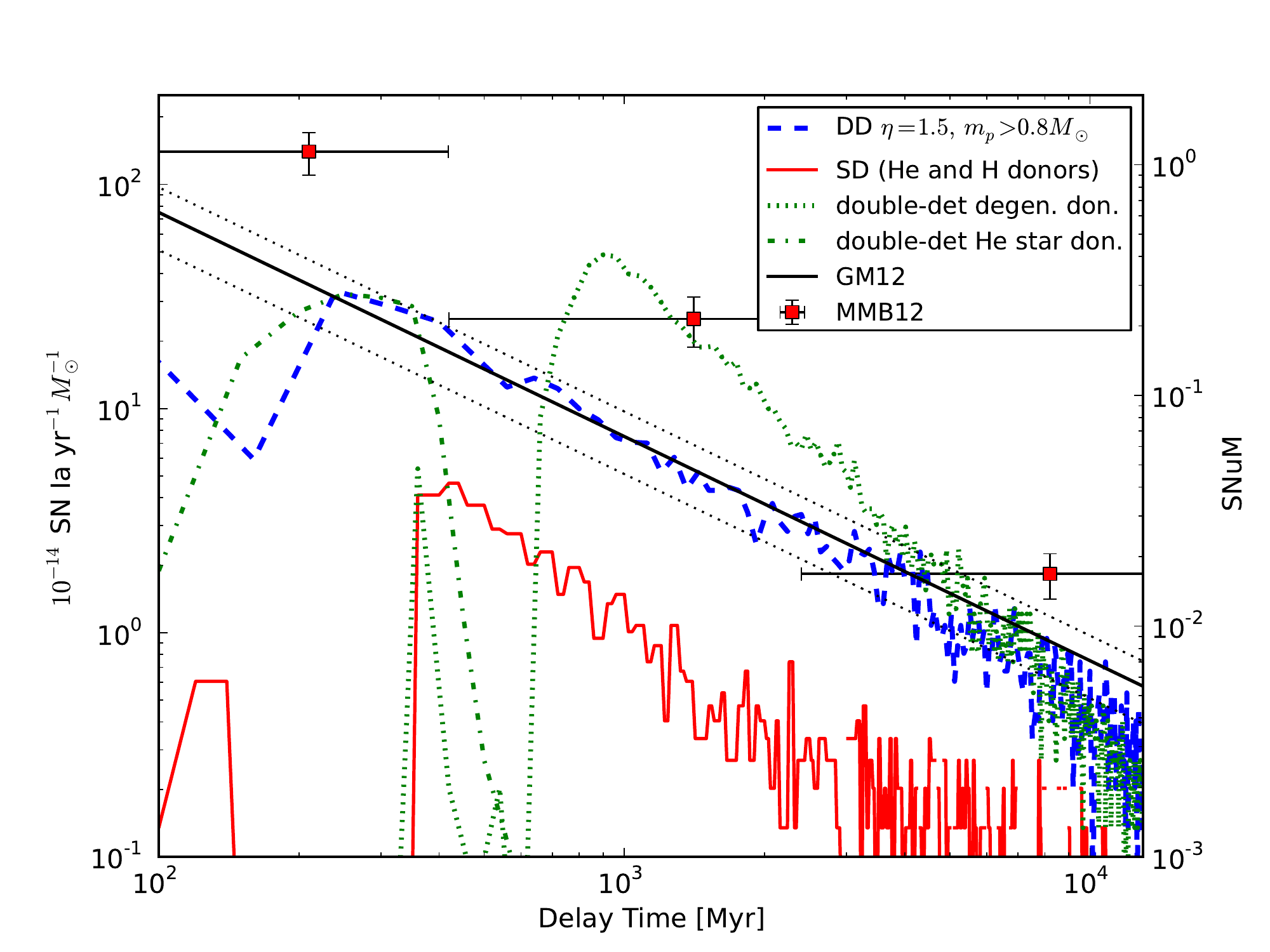}
  \caption{Coloured lines show {\sc StarTrack} 
    theoretical delay time distributions
    from $100 - 13000$ Myr after star formation assuming a $70$ \%
    binary fraction.  The DTDs have been mass-normalized into units of
    SNuM on the right-hand y-axis for comparison with observations  
    (SNe Ia ($10^{10}$ \msun)$^{-1}$ ($100$ yr)$^{-1}$). 
    The lines have been smoothed,
    and small fluctuations are due to Monte Carlo noise.  
    Dashed blue: DD violent WD mergers, which are a DD scenario 
    sub-class (cf.~Sect.~\ref{sec:assess}); 
    solid red: SD Chandrasekhar-mass scenario including main sequence, 
    giant-like 
    and helium-burning donors; green dotted: double-detonation
    scenario (see also \citep{ruiter2011a}) with helium-rich 
    WD donors; green dash-dotted:
    double-detonation scenario with helium-burning star donors.  
    Alongside the model DTDs, we show the recent 
    observationally-recovered
    DTD from \citet{maoz2012a}(red squares) and the DTD
    from observations as described in \citet{graur2012a} (straight 
    black lines; $t^{-1}$ best-fit).}
  \label{fig:obs-DTD}
\end{figure}

With binary population synthesis models, the entire evolutionary
history of each binary is followed, so the DTD is easily determined
for all potential SN Ia progenitors.  In Figure \ref{fig:obs-DTD} we
show mass-normalized theoretical DTDs for the following progenitor
scenarios: SD (hydrogen-burning and helium-burning donors); a
sub-class of the DD, whereby the mass of the primary WD must be $\ge
0.8$ \msun and there are additional restrictions on the mass ratio,
(cf.~Sect.\ \ref{sec:violent_mergers}); and double-detonation
progenitors involving both helium-burning star and helium-rich WD
donors \citep{ruiter2009a,ruiter2011a,ruiter2012a}.  Alongside our
theoretical DTDs we show the most recent DTDs from (red squares,
\citet{maoz2012a}) and the best-fit DTD from observations as described
in \citet{graur2012a}($t^{-1}$ power-law best-fit). All of the DTDs
are normalized per mass formed in stars, and can be thought of as an
absolute SN Ia rate as a function of Hubble time.  We note that the
observational DTDs, while currently the most recently-derived, have
amplitudes that are at least a factor of a few lower than cosmic DTDs
derived from previous studies (see \citet{maoz2012a} for discussion).

It is immediately obvious that no single progenitor channel can
reproduce the observed rates for delay times $< 300$ Myr.  The SD
(red, solid line) `spike' at delay times $<200$ Myr is solely due to
helium-rich donors; these donors have relatively faster evolutionary
timescales than their hydrogen-donor SD counterparts (hence the gap
$\sim 200-300$ Myr).  The SD DTD drops off too quickly to follow a
$t^{-1}$ power-law, but shows the expected trend at delay times $>400$
Myr (decreasing events with increasing delay time). Overall, the SD
rates are too low by about an order of magnitude to match the observed
SN Ia rate.

The double-detonation progenitors display a clear bimodal behaviour:
prompt events originate from systems with helium-burning star donors
(green dash-dotted line; short evolutionary timescales) while delayed
events originate from systems where the donor is a degenerate dwarf
(green dotted line; longer evolutionary timescales).  Beyond $\sim
300$ Myr, the rates match the observations fairly well, and beyond 1
Gyr the DTD follows a (steeper) power law $\sim t^{-2}$ (see
  also \citep{ruiter2011a}).  However, there is still a lot of uncertainty
with respect to the explosion mechanism of this channel regarding the
detonation of the helium shell (Sect.~\ref{sec:sub_MCh}).

The sub-set of DD mergers shown in Figure \ref{fig:obs-DTD} -- a
population of violent mergers (blue dashed line) -- are discussed in
detail in \citet{ruiter2012a} (see also
Sect.~\ref{sec:violent_mergers}).  Once a double WD is born, it may
take several Gyr before the stars are brought into contact.  The
dominant mechanism leading to a decrease in orbital angular momentum
(and hence smaller orbit) is the emission of gravitational waves
\citep{peters1964a}.  As is expected, the DD mergers shown here follow
the $t^{-1}$ power-law shape fairly well, and are within the
observational uncertainties for delay times $> 300$ Myr (below $300$
Myr there is a dip in the distribution, see Sect.~\ref{sec:assess} for
explanation).  It may also be possible to increase the overall rates
of DD mergers if three-body interactions are taken into account
\citep{thompson2011a}.

\section{Modeling explosion and formation of observables}
\label{sec:models}

\subsection{The MPA modeling pipeline}

A thorough testing of different SN~Ia scenarios requires to combine
several model aspects from progenitor evolution over the explosion to the
formation of the observables. A way of addressing the progenitor
problem was discussed in Sect.~\ref{sec:progenitors}. Here we present
our approaches to modeling hydrodynamics and nucleosynthesis in the
thermonuclear explosion and radiative transfer in the ejecta cloud
giving rise to the observables. In the subsections which follow we
first describe our combustion-hydrodynamic code \textsc{leafs} and then
the Monte-Carlo radiative-transfer code \textsc{artis}.

\subsubsection{Hydrodynamic explosion models}
\label{sec:models_hydro}

Type Ia supernova explosion models aim at following the hydrodynamical
evolution from the ignition of thermonuclear burning in a WD
to homologous expansion of the ejecta. These models rely on the
equations of hydrodynamics (either the Euler equations or specific
approximations suitable for low-Mach number flows) coupled to
nuclear reactions. After ignition, a combustion wave forms. Because of
the high temperature sensitivity of carbon fusion, it is
confined to a narrow region in space. Seen from the scales of the
WD, it can be approximated as a sharp discontinuity
separating the fuel (CO material) from the ashes of the
nuclear burning. The reactive Euler equations allow (in their integral
form) for two distinct classes of discontinuous (weak) solutions that
model the propagation of such thin combustion waves: subsonic
\emph{deflagrations} and supersonic
\emph{detonations}. Microscopically, the deflagration propagation mode
corresponds to a flame mediated by heat conduction, while a detonation
is driven by a shock wave.

The most critical aspect in modeling thermonuclear explosions of 
WDs are the inherent scale problems -- both in time and
space. Although in large-scale simulations of supernova explosions
that capture the entire WD star it is well justified to treat
the combustion fronts as discontinuities (at least for most of the
burning taking place at high fuel densities), this implies that the
internal structure of the flame cannot be resolved. Thus, details of
their mechanism and in particular
the nuclear reactions are not represented. A time scale problem is the
discrepancy between the scales of hydrodynamic flows and that of the
nuclear reactions. In Chandrasekhar-mass models, the ignition of the
combustion wave is preceded by a period of convective carbon burning
(the so-called ``simmering phase'') that lasts for a century and is
characterized by highly turbulent flows. A correct modeling of
turbulence is also essential for deflagration phases in
Chandrasekhar-mass explosion models. Here, Reynolds numbers of the
order of $10^{14}$ are typically encountered posing another severe
spatial scale problem.

While a detailed modeling of the ignition process and the simmering
phase in Chandrasekhar-mass explosion models remains challenging (but
see \citep{zingale2009a,nonaka2012a} for recent efforts), there have
been several attempts to overcome the spatial scale problem associated
with combustion fronts. In large-scale supernova simulations, an
option is to artificially broaden the combustion waves so that they
can be represented on the computational grid. This flame-capturing
approach was explored by \citet{khokhlov1993a,khokhlov1995a}. The main
drawback is that an artificially-broadened flame smears out
small-scale dynamics. Level-set based techniques for combustion wave
tracking are an alternative which has been introduced to SN~Ia
simulations by \citet{reinecke1999a,reinecke1999b}. The underlying
idea is to represent a contour (in 2D simulations) or a surface (in 3D
simulations) by the zero-level set of a scalar field that is set up as
a signed distance function to the combustion wave. This scalar field
is then evolved in an appropriate way to model the propagation of the
burning front \citep{reinecke1999a}. In this approach, the combustion
wave is considered as a sharp discontinuity and no attempt is made to
resolve its inner structure. It therefore has to be augmented by a
model for the propagation speed of the combustion wave and the energy
release in it. For deflagrations, the laminar flame speed has been
determined in small-scale simulations \citep{timmes1992a}. This sets
the lower limit of the propagation speed of the effective flame front
in large-scale supernova simulations. For most of the time, turbulence
determines the flame speed and this is accounted for by a specific
modeling approach (see below). For detonations, the propagation speed
in the most simple case is the Chapman-Jouget speed, i.e.\ the sound
speed in the ashes. This, however, does not take into account the
dependence on shock strength and possible multi-dimensional
effects. In addition, at high densities burning to nuclear statistical
equilibrium makes part of the process endothermic and pathological
detonations are encountered here. For these, the microscopic mechanism
has been studied by \citet{sharpe1999a} and the results can be used to
model the propagation of detonations in large-scale supernova
simulations.

The available computational resources limit the reaction
networks employed in the hydrodynamic explosion simulations to only a
few species. This allows to represent the energy release to a
sufficient precision to account for its impact on the dynamics of the
explosion. The detailed nucleosynthetic yields, however, cannot be
determined directly. These are required to compare the
models with the constraints from galactic chemical evolution and for
setting the input models for radiative transfer simulations that rely
on a spatially resolved multi-dimensional chemical structure of the
ejecta. This problem is usually accounted for by a nucleosynthetic
postprocessing step \citep{travaglio2004a,seitenzahl2010a}. In the 
hydrodynamic explosion model, a large
number (up to several million) of Lagrangian tracer particles are
advected with the flow. They represent fluid packages and for these
the thermodynamic trajectories are recorded. These data are then used as
input for a large reaction network that allows to reconstruct the
details of the nuclear reactions. 

Finally, the expansion of the WD and the ejecta in the course of the
explosion pose another scale problem. With ejecta velocities well
above $20,000 \, \mathrm{km}\,\mathrm{s}^{-1}$ the material would
quickly leave a static computational grid. Two approaches to overcome
this problem are followed. Adaptive mesh refinement allows to use
large computational domains with finer resolution at places where
physical processes are to be resolved. An alternative, which is
implemented in our models, is to use moving computational meshes 
\citep{roepke2005c,roepke2005a,roepke2006a}. As
the overall expansion of the ejecta is spherical to first order, a
simple radial expansion of the computational grid provides an optimal
resolution of the explosion physics with given computational resources
and allows for following the hydrodynamic evolution of the ejecta to a
relaxed state (homologous expansion) -- a prerequisite for predicting
observables with radiative transfer simulations.

\subsubsection{Radiative transfer models}

>From the hydrodynamic explosion simulations we obtain the velocities, 
densities and composition of the explosion ejecta.  These, however,
are not directly comparable to the observational signatures of SNe~Ia
like broad-band photometry, spectral time series and spectropolarimetry
over a wide range of the electro-magnetic spectrum.  For that purpose
synthetic spectra and light curves must be obtained from radiative
transfer calculations.  Since the explosion ejecta are free streaming
at about 100\,s after the explosion (e.g. \citep{roepke2005c}), the 
radiative transfer calculations can be decoupled from the hydrodynamic 
simulation assuming homologous expansion of the ejecta.

The observational display of SNe~Ia is not powered by the heat produced
during the explosion itself.  Already at the first observational epochs, 
typically a few hours to days after the explosion, this heat has long 
gone due to the expansion of the ejecta.  Instead, the decays of 
radioactive isotopes like \nuc{56}{Ni} and \nuc{56}{Co}, freshly 
synthesized during the thermonuclear burning, give rise to the 
emission of a spectrum of $\gamma$-photons.  These interact with the 
ejecta by Compton scattering, pair production and photoelectric 
absorption, thereby depositing their energy and reheating the ejecta
\citep{truran1967a,colgate1969a}.  Thus, radiative transfer simulations 
that aim at a direct connection between explosion models and 
observations have to take into account this energy injection and the 
transport of $\gamma$-photons explicitly.  A simple photospheric
assumption (e.g. \citep{mazzali1993a}) is not enough.

Another complication poses the peculiar chemical composition of
SNe~Ia.  Since their ejecta do not contain any hydrogen but
significant amounts of iron-group elements, the opacity in SN~Ia is
dominated by the wealth of lines associated with the iron-group
elements (e.g. \citep{pinto2000b}, figure~1), thus requiring a
solution of the complicated multi-line transfer problem in expanding
media.  Assuming emission from a photosphere and spherical symmetry,
many studies have addressed this problem in the past either assuming
pure resonance scattering
(e.g. \citep{branch1982a,branch1983a,mazzali1993a}) or pure
absorption (e.g. \citep{jeffery1992a}) in the lines.  However, such
an approach is too simple, since it cannot account for line
fluorescence effects which are crucial in shaping the spectral energy
distribution of SNe~Ia \citep{lucy1999b,pinto2000b}.

Finally, given the complex ejecta structure of state-of-the art 
hydrodynamic explosion models a time-dependent 3D ``full-star'' 
treatment of radiative transfer which simulates the $\gamma$-deposition 
and spectrum formation in detail is needed.  Such an approach is 
e.g.\ taken in the Monte Carlo radiative transfer code {\sc sedona}
of \citet{kasen2006a} which treats line fluorescence in an approximate 
way similar to \cite{lucy1999b,pinto2000b}.  

Following the methods outlined by \citet{lucy2002a,lucy2003a,lucy2005a}, 
at MPA we have developed another time-dependent 3D Monte Carlo 
radiative transfer code {\sc artis} \citep{kromer2009a,sim2007b}.
{\sc artis} divides the total energy available in the radioactive 
isotopes of a given supernova model into discrete energy packets. 
These are initially placed on a computational grid according 
to the distribution of the radioactive isotopes and then follow 
the homologous expansion until they decay.  Upon decay they convert 
to bundles of monochromatic $\gamma$-ray photons which propagate 
through the ejecta.  {\sc artis} contains a detailed treatment 
of $\gamma$-ray radiative transfer \citep{sim2008a} and accounts 
for interactions of $\gamma$-ray photons with matter by Compton 
scattering, photo-electric absorption and pair production.
Assuming instantaneous thermalization of absorbed $\gamma$-ray 
photons, the energy is transformed into ultraviolet-optical-infrared 
photons enforcing statistical and thermal equilibrium. Using a 
detailed wavelength-dependent opacity treatment, {\sc artis} solves 
the radiative transfer problem self-consistently with the ionization 
and thermal balance equations.  Excitation is treated approximately 
by assuming local thermodynamic equilibrium, which is expected to be 
a good approximation at least around maximum light.  A generalized 
treatment of line formation \citep{lucy2002a,lucy2003a}, including 
typically about 500,000 individual atomic line transitions 
\citep{kurucz1995a} in the Sobolev approximation \citep{sobolev1957a}, 
allows for a detailed treatment of radiation-matter interactions 
including a parameter-free treatment of line fluorescence. Thus, 
depending only on the input model and atomic data, our radiative 
transfer calculations give a maximum of predictive power for a 
given explosion model.

\subsection{Models for normal SNe~Ia}
\label{sec:normals}

As discussed in Sect.~\ref{sec:observations}, normal SNe~Ia can be
explained by the decay of typically 0.3\,\msun\ to 0.9\,\msun
of $^{56}$Ni in the center of the ejecta which is surrounded by layers
of intermediate mass elements, oxygen and unburnt material. There are
several ways of constructing explosion models that give rise to such
an ejecta structure.

\subsubsection{Chandrasekhar-mass delayed detonations}
\label{sec:normals_from_MCh}

The model of a WD exploding when approaching the
Chandrasekhar mass is certainly the most thoroughly explored
option. Traditionally, it is associated with the single-degenerate
progenitor model (see Sect.~\ref{sec:progenitors}), but the formation 
of a Chandrasekhar-mass 
WD due to a merger is not excluded. A strong argument in favor of
the Chandrasekhar-mass model was the notion of homogeneity among
SNe~Ia. This picture, however, eroded with the detailed observational
campaigns of the past decade (see Sect.~\ref{sec:observations}) that
clearly showed a pronounced diversity among these objects. By now,
several sub-classes have been established. Therefore, it seems
unlikely that all SNe~Ia can be explained within a single
progenitor/explosion model. Nonetheless, from the explosion modeling
point of view, the Chandrasekhar-mass scenario holds promise to
explain the bulk of normal SNe~Ia. There may, however, be difficulties
with explaining the rate of observed events, when this scenario arises
exclusively in the single-degenerate channel (see 
Sect.~\ref{sec:progenitors}).

One-dimensional parametrized models of explosions in
Chandrasekhar-mass WDs have been very successful in reproducing normal
SNe~Ia, most notably the W7 model of \citet{nomoto1984a}. Here, we will
focus on recent developments in simulating the nuclear burning in
two or three spatial dimensions. There is a qualitative difference between
such models and earlier one-dimensional parameterizations. On the one
hand, multidimensional approaches allow for a more realistic
treatment of  inherently multidimensional effects such as turbulent
burning and asymmetries in the ignition and flame propagation. On the
other hand, by fixing free parameters, such models cannot easily be
used to fit observations and thus the level of agreement with
observations is usually lower. Thus, the
assessment on the validity of the underlying models is more
involved and interpretation is required.

The first numerically studied explosion model, a  
\emph{prompt detonation of a Chandrasekhar-mass WD}
\citep{arnett1969a} can be ruled out as an explanation for SNe~Ia. 
Since detonations propagate at supersonic
velocities with respect to the fuel, there is no causal contact
between the energy release and the material ahead of the combustion
wave. Thus, the entire star burns at the high initial densities (a
few times 10$^9$ g/cm$^3$) of a
Chandrasekhar-mass WD in hydrostatic equilibrium. Consequently, burning 
proceeds to nuclear statistical equilibrium (complete burning)
throughout most parts of the star and
the ejecta consist almost exclusively of iron group elements
(predominantly $^{56}$Ni). This is in conflict with the observational
requirements for normal SNe~Ia, see Sect.~\ref{sec:observations}. In 
order to produce less $^{56}$Ni and
a substantial amount of intermediate mass elements, at least parts of
the burning must proceed at lower densities than those encountered in
Chandrasekhar-mass WDs in hydrostatic equilibrium. 

A combustion starting out in the deflagration mode brings the WD out
of equilibrium and pre-expands the fuel material. Consequently,
burning partially takes place at lower densities than in an
equilibrium Chandrasekhar-mass WD. This allows for the synthesis of
intermediate-mass elements and reduces the $^{56}$Ni yield
accordingly. However, ultimately laminar deflagration flames are too
slow to catch up with the expansion of the star. This limits the
amount of material burnt and thus the nuclear energy release is too
low for a successful SN~Ia. It has been noted early on
\citep{nomoto1984a} that deflagrations will not propagate at their
laminar speeds. Burning from the WD's center outward, they produce an
inverse density stratification in the gravitational field of the star
and it is thus subject to buoyancy instabilities. The Rayleigh-Taylor
instability and secondary shear instabilities generate strong
turbulence. Driven from large scales, the turbulent energy cascades
down to the microscopic Kolmogorov scale. Consequently, the flame
interacts with turbulent eddies of various sizes. The flame is torn
and wrinkled by these turbulent motions and this enlargement of the
flame surface area accelerates its mean propagation significantly.

Several numerical studies indicate that although turbulence is driven
on large scales by buoyancy, it quickly becomes isotropic and follows
Kolmogorov-scaling at smaller scales
\citep{zingale2005a,ciaraldi2009a}. The correct representation of
flame-turbulence interaction is one of the key challenges in modeling
deflagrations in WDs and thus a critical ingredient in
Chandrasekhar-mass models for SNe~Ia. Several possibilities have been
suggested to accomplish this. In the work discussed here, a
subgrid-scale model is employed. It is based on a balance equation for
the unresolved turbulent kinetic energy. For two-dimensional
simulations, the approach of \citet{niemeyer1995b} is used while
three-dimensional simulations use the method of
\citet{schmidt2006c} that does not make any assumptions on the scaling
of turbulence. For strong turbulence, as expected for most
phases of the supernova explosion, flame-turbulence interaction
implies that on some sufficiently large scale (such as resolved in
multi-dimensional simulations) the propagation speed of the effective
flame (averaged over unresolved small-scale structure) decouples from
the laminar burning speed and is set by the turbulent velocity
fluctuations on that scale \citep{damkoehler1940a}. This is the basis
for the flame model in our simulations that employ the level-set
technique to represent the effective deflagration front and use a
subgrid-scale turbulence model for determining its effective
propagation velocity (for details see also \citep{roepke2009a}). 

The amount of burning and the energy release depend strongly on the
way the flame is ignited. In Chandrasekhar-mass explosions a century
of convective carbon burning precedes the actual flame ignition (cf.~
Sect.~\ref{sec:models_hydro}). Numerical simulations of this phase are
extremely challenging due to its long duration and the high turbulence
intensities involved (but see \citep{hoeflich2002a,kuhlen2006a,
  zingale2009a} for recent attempts). At the moment, the geometry of
flame ignition is unclear and therefore different possibilities are
considered. If ignited in many sparks around the center, the WD can be
unbound \citep{roepke2006a, roepke2007c}.  But even with a strong
ignition, the asymptotic kinetic energy of the ejecta does not exceed
$\sim$$0.6 \times 10^{51} \, \mathrm{erg}$ and the $^{56}$Ni
production reaches at best about a third of a solar mass
\citep{roepke2007c}. The most optimistic values for pure deflagrations
in Chandrasekhar-mass WDs reach the fainter end of \emph{normal}
SNe~Ia, but they cannot account for all of them.  Moreover, the
predicted spectra show peculiarities that can be attributed to a
chemically mixed ejecta composition which is a natural consequence of
the large-scale buoyancy instabilities in these models. Thus we
conclude that deflagrations in Chandrasekhar-mass WDs cannot explain
\emph{normal} SNe~Ia. They could, however, account for a peculiar
subclass (see Sect. \ref{sec:2002cx-likes}).

\begin{figure*}
  \centering
  \includegraphics[width=0.329\linewidth]{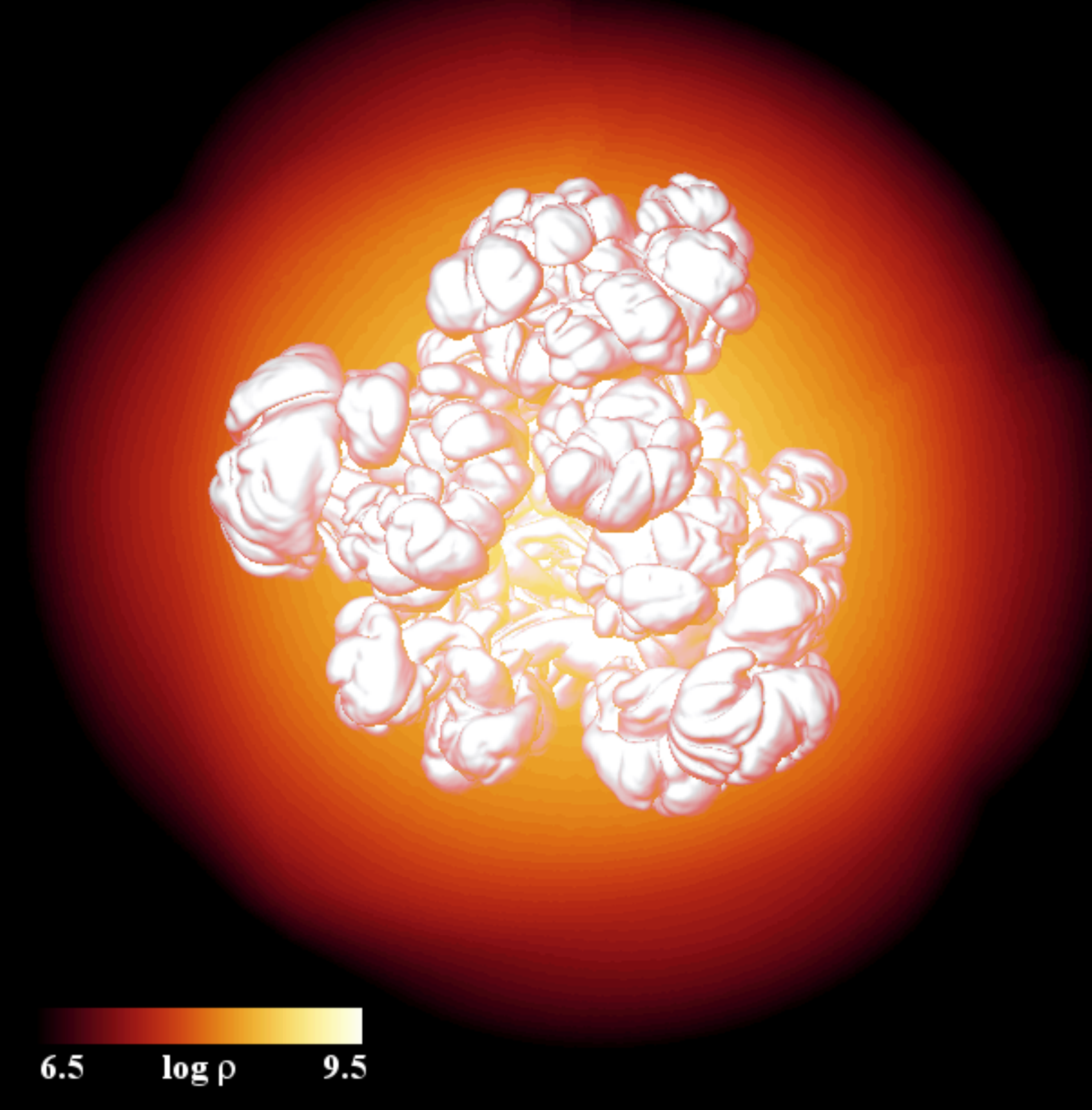}
  \includegraphics[width=0.329\linewidth]{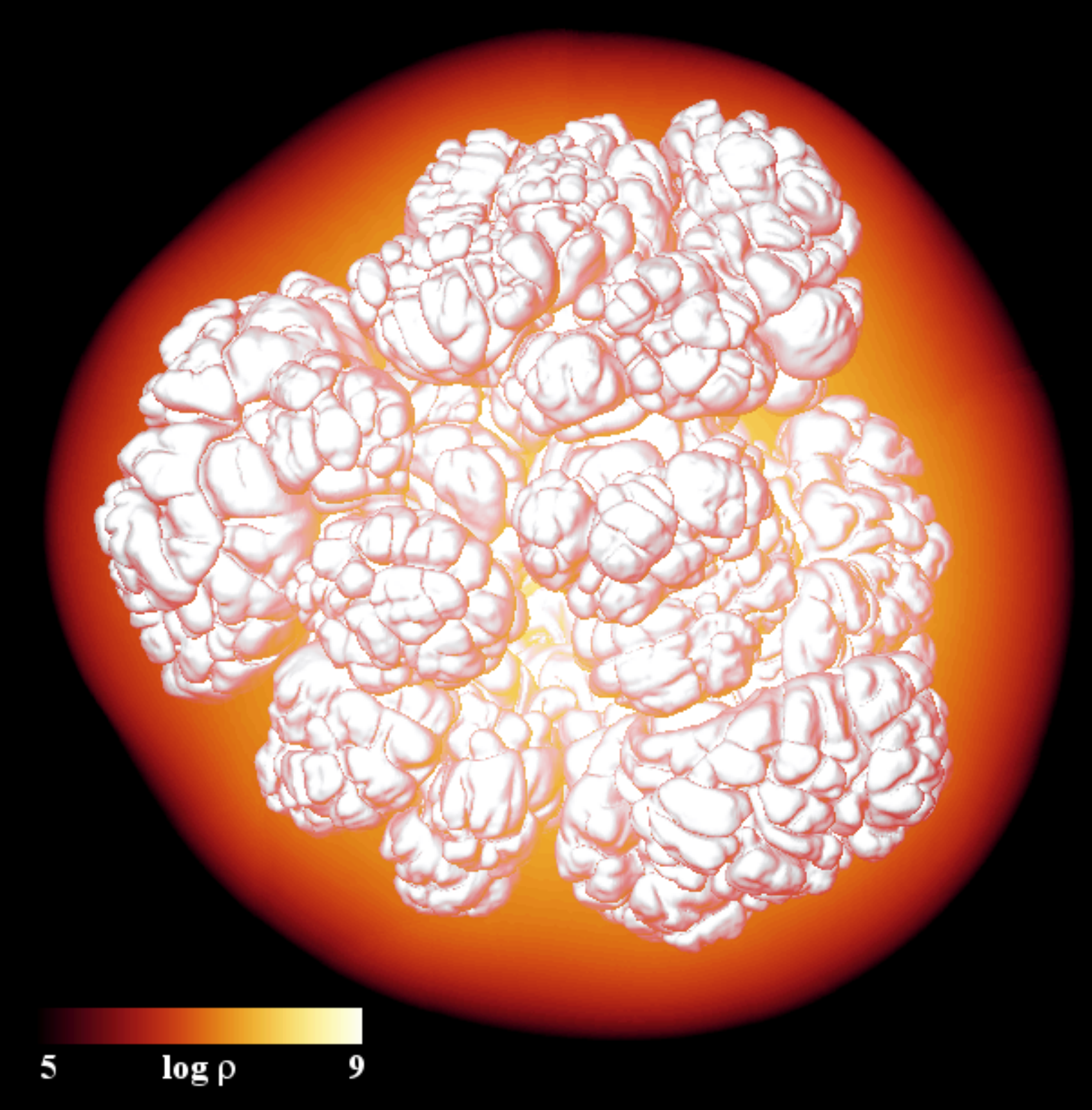}
  \includegraphics[width=0.329\linewidth]{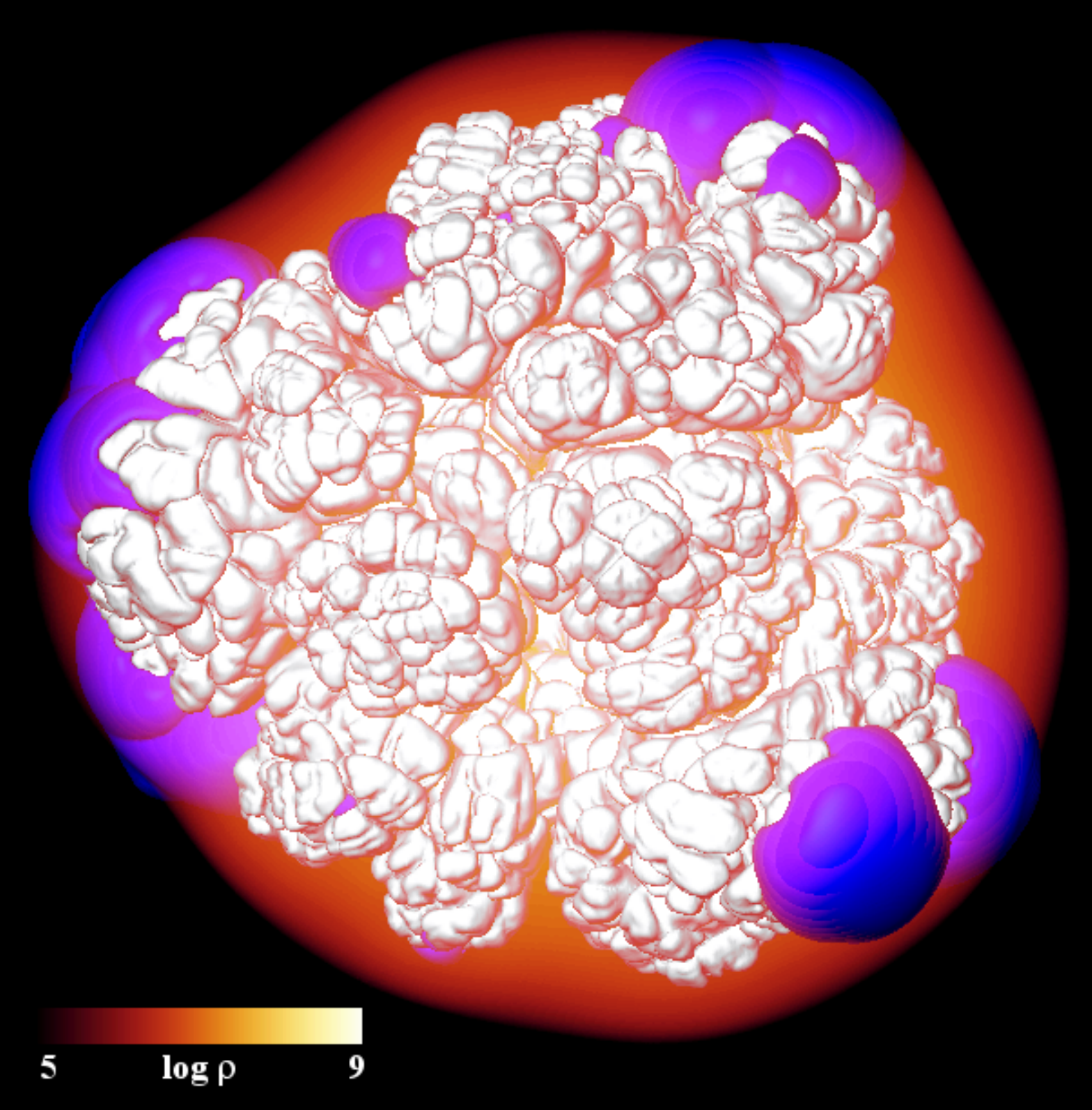}
  \caption{Hydrodynamic evolution of a Chandrasekhar-mass delayed
    detonation. Shown are a volume rendering of the density
    (orange colours) and the zero level-set of the deflagration
    (whitish surface) and detonation flames (blueish colours)
    of model N100 \citep{roepke2012a, seitenzahl2012a}. From left to
    right the snapshots are taken at
    0.70, 0.93 and 1.00\,s.}
  \label{fig:n100_deldet}
\end{figure*}

The only chance for Chandrasekhar-mass explosion models to reach the
ballpark of normal SNe~Ia is a detonation following the initial
burning in the deflagration mode. In contrast to a prompt detonation
of a Chandrasekhar-mass WD in hydrostatic equilibrium, the
pre-expansion in the deflagration phase now allows the detonation to
burn at lower fuel densities. Although it still can contribute to the
overall $^{56}$Ni production, it produces a substantial layer of
intermediate mass elements in the outer layers of the exploding
WD. One way to realize this is the \emph{delayed-detonation scenario}
\citep{khokhlov1991a}, in which a spontaneous transition of the
burning front from deflagration to detonation occurs in a late stage
of the explosion. This leads to a clear chemical stratification with
iron group elements dominating the inner part of the ejecta while the
products of a detonation in material of subsequently lower density
lead to a stratified composition in the outer layers. Here,
intermediate-mass elements follow the iron group elements and at
higher velocities oxygen and carbon dominate.  Downdrafts of unburned
material left behind in the turbulent and unstable deflagration are
now incinerated. A qualitative difference to earlier one-dimensional
delayed-detonation models, however, is that stable iron group elements
produced in the high-density deflagration at the center of the WD do
not stay there but float to larger radii due to buoyancy
instabilities.  The degree of the pre-expansion and thus the total
$^{56}$Ni production is determined by the energy release in the
deflagration \citep{roepke2007b, mazzali2007a} and by the delay
between deflagration ignition and detonation triggering. One way to
vary the strength of the deflagration is by choosing different ignition
configurations (although other parameters may also affect the
strength of the deflagration phase, see e.g. \citep{roepke2006b,
krueger2010a, jackson2010a, seitenzahl2011a}). Igniting vigorously in
many ignition sparks around the WD center (e.g., \citep{roepke2006a,
  schmidt2006a}) releases more energy in the deflagration burning,
hence achieving more pre-expansion \citep{roepke2007b, mazzali2007a},
while a sparse and perhaps asymmetric ignition leads to a weak
deflagration phase (e.g., \citep{roepke2007a, jordan2008a}). In the
context of the delayed-detonation explosion scenario this gives rise
to a variability of $^{56}$Ni production which, in turn, leads to a
range in brightnesses of the simulated events covering that of
\emph{normal} SNe~Ia. The brightness of the faintest model is set by
the strongest pre-expansion and thus by the most vigorous deflagration
that is achievable. For nearly isotropic ignitions with standard WD
setups this corresponds to a $^{56}$Ni production in the range of
$[0.3\ldots0.4]$\,\msun\ -- clearly too much for subluminous
SNe~Ia. On the other end, weak deflagrations arising from asymmetric
ignitions easily lead to the production of up to a solar mass of
$^{56}$Ni in the delayed-detonation scenario. Thus, in principle, this
model should be able to reproduce the range of observed brightnesses
of normal SNe~Ia.

\begin{figure*}
  \centering
  \includegraphics[width=0.329\linewidth]{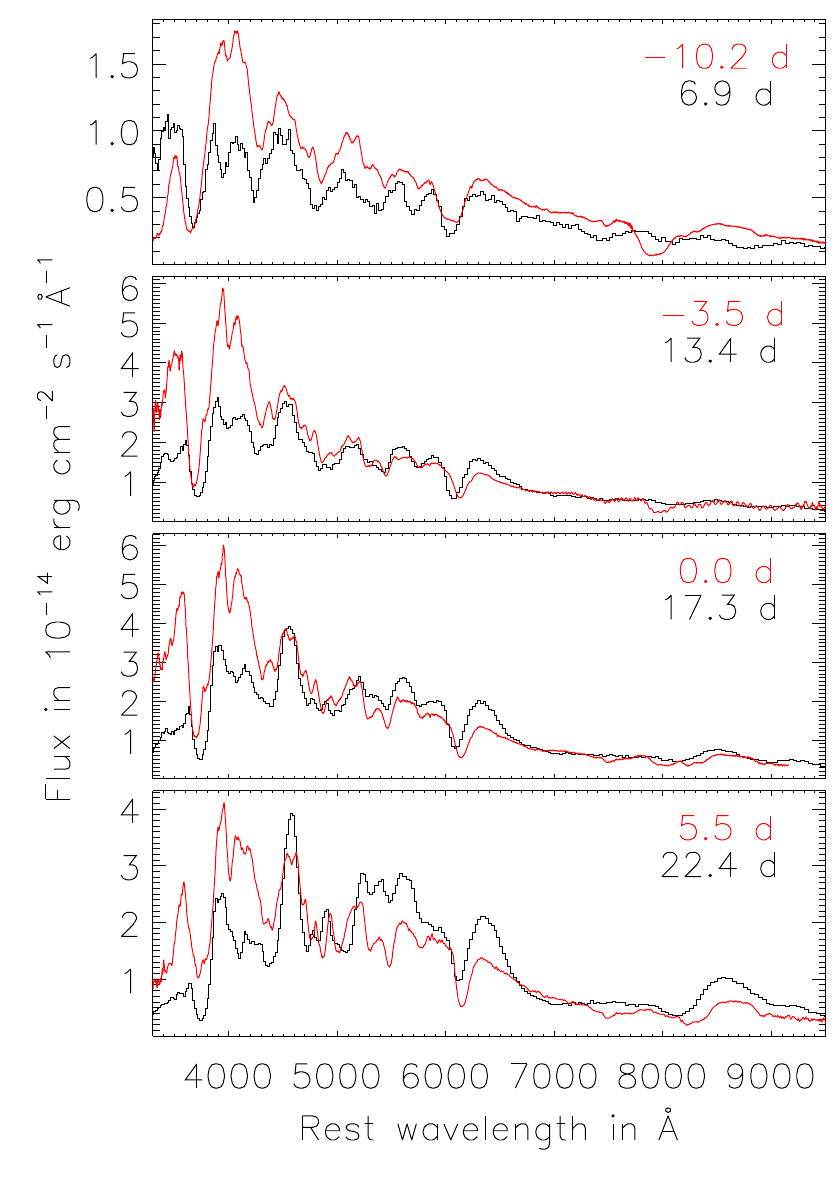}
  \includegraphics[width=0.329\linewidth]{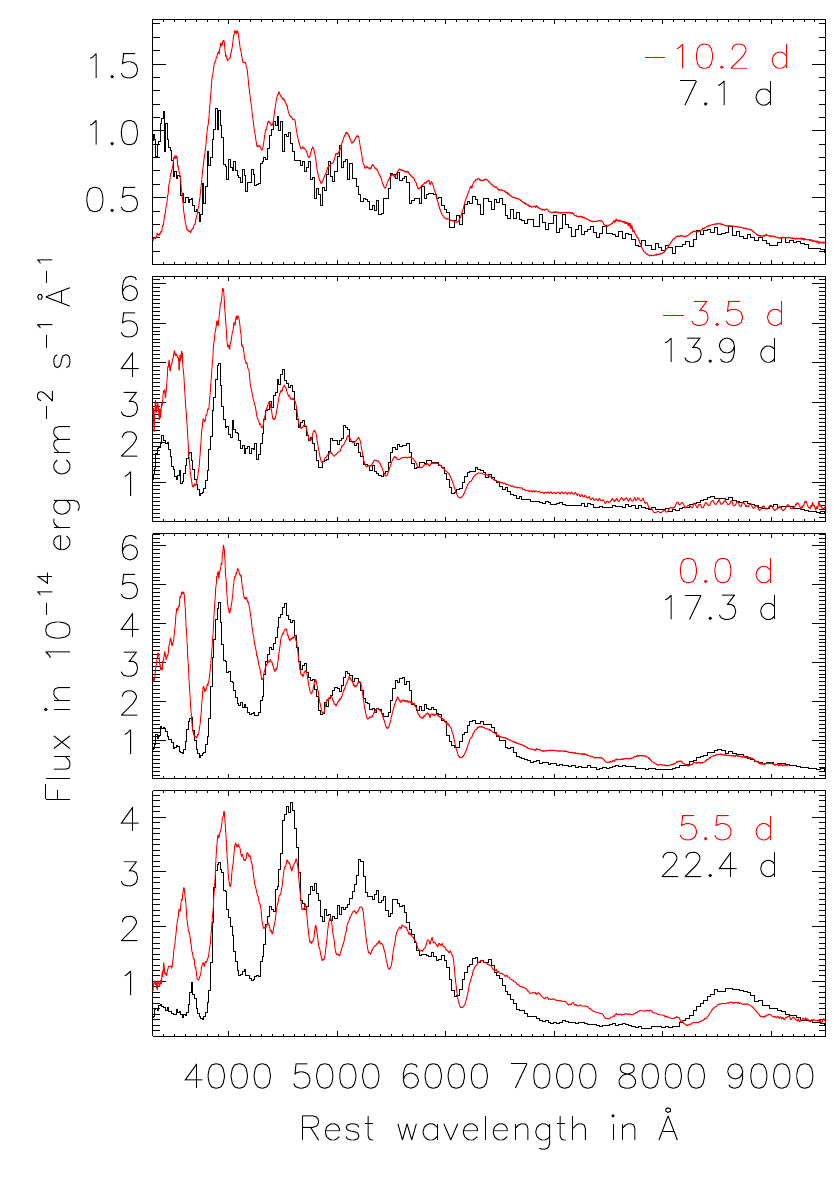}
  \includegraphics[width=0.329\linewidth]{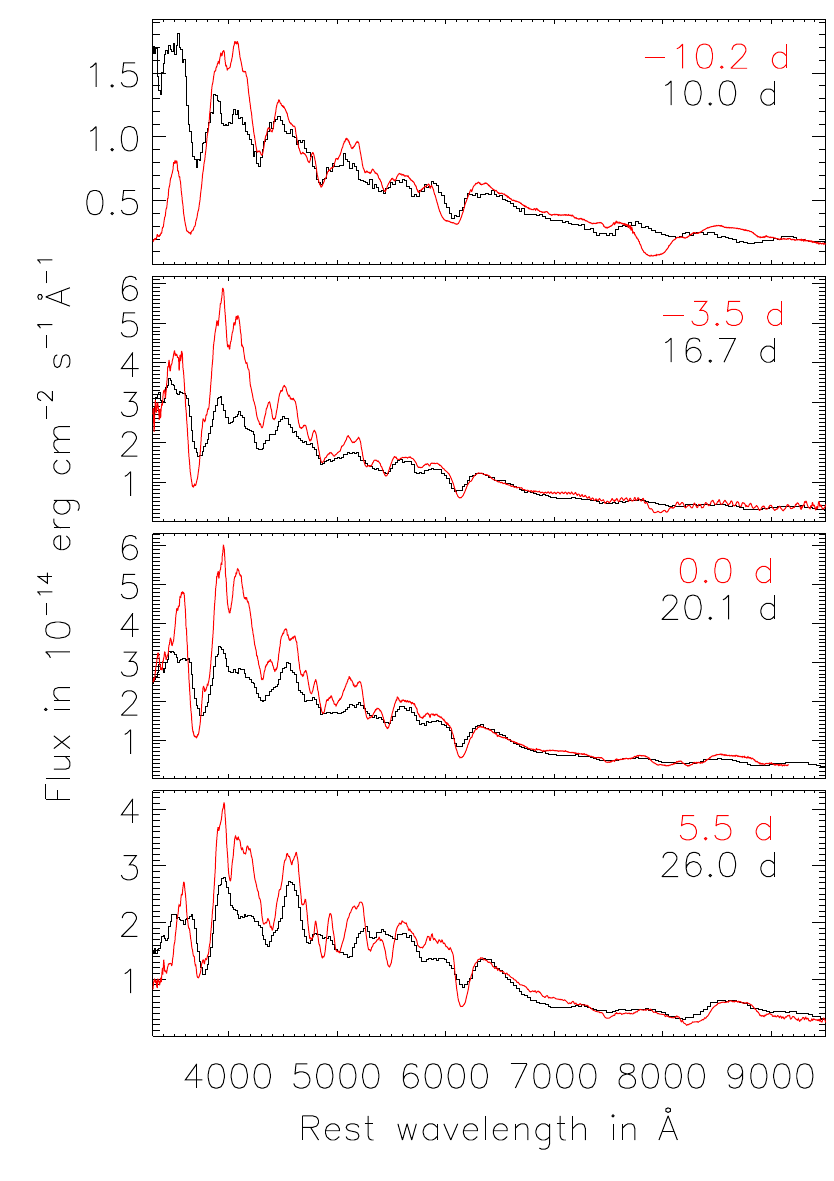}
  \caption{Synthetic spectra of different kinds of explosion 
    models for normal SNe~Ia. From left to right the panels show 
    (i) the delayed-detonation model N100 \citep{roepke2012a}, 
    (ii) model 3m of the sub-Chandrasekhar-mass double detonations
    presented by \citet{kromer2010a}, and (iii) a double-degenerate 
    merger of two WDs with 1.1 and 0.9\,\msun\ \citep{pakmor2012a}. 
    For comparison, we show observed spectra of the ``golden-standard'' 
    normal SN~Ia 2005cf for corresponding epochs \citep{garavini2007a}
    (data in red).}
  \label{fig:specplot_normals}
\end{figure*}

As an example, Fig.~\ref{fig:n100_deldet} shows model N100
\citep{roepke2012a,seitenzahl2012a} which is ignited in 100 ignition
sparks around the center. The ensuing deflagration (left panel) is of
intermediate strength. The middle panel shows the deflagration front
directly prior to the first deflagration-to-detonation transition. The
large-scale buoyancy-induced plumes of burnt material are clearly
visible. This -- together with shear-induced turbulence on smaller
scales leads to the increase in flame surface area characteristic for
the turbulent deflagration. The panel on the right hand side shows a
snapshot shortly after the first deflagration-to-detonation transition
has triggered. Obviously, it is immediately followed by other
transitions at different locations. The newly formed detonation waves
quickly spread over the remaining fuel and burn out the downdrafts of
fuel material left behind from the deflagration. Since the detonation
propagates from high to low density the ash composition changes from
iron-group to intermediate-mass nuclei and, because of the supersonic
propagation, there is no mixing, in contrast to the deflagration
phase. The outcome is an ejecta cloud with a stratified chemical
composition in the outer layers and close to 0.6\,\msun\ of $^{56}$Ni
at the center. The hydrodynamic evolution is followed with a
moving-grid technique to $100\,\mathrm{s}$ after ignition. After
nucleosynthetic postprocessing, the ejecta structure is mapped into
the radiative transfer code \textsc{artis} \citep{kromer2009a,
  sim2007b} to calculate synthetic observables. A sequence of spectra
for this model is shown in the left panel of
Fig.~\ref{fig:specplot_normals}. Overall, the agreement between the
model spectra and the observational reference spectra of a normal
SN~Ia (SN~2005cf) is reasonable. Again, we emphasize that no perfect
match is expected in this comparison of a generic three-dimensional
supernova model and an observation without any attempts of
fitting. However, a more fundamental shortcoming of the model is that
it appears to be too red. This can be attributed to a flux
redistribution due to stable iron group elements at rather high
velocities -- a feature that at least to a certain degree is
characteristic for delayed-detonation models.

A more systematic test has been presented by \citet{kasen2009a} on the
basis of a suite of two-dimensional models. Again,
although no perfect agreement with observational data is reached, many
of the models would be classified as SNe~Ia employing a tool for
analyzing observations and treating the models as actual astronomical data
\citep{blondin2011a}. However, the brightest and most asymmetric
explosions in the \citealt{kasen2009a} sample would not be classified
as SNe~Ia. Interestingly, in this set of models, 
the correlation between peak luminosity in the $B$-band and the
decline rate of the light curve (used to calibrate SNe~Ia as distance
indicators in observational cosmology, \citep{phillips1993a,phillips1999a})
was found to resemble that of the observations
\citep{kasen2009a}. Whether or not this is the case also in sets of
three-dimensional models remains to be seen and is subject to forthcoming
publications (Sim et al., in preparation).

\subsubsection{Sub-Chandrasekhar-mass double detonations}
\label{sec:sub_MCh}

The observational finding of chemically stratified ejecta points to a
detonation propagating down the gradient towards low densities in the
outer layers of the exploding WD. As discussed above, for
Chandrasekhar-mass WDs this is only compatible with a configuration
that is out of hydrostatic equilibrium. An alternative to this
mechanism is a detonation in a sub-Chandrasekhar mass WD. Pure
detonations in CO WDs with masses between 0.81\,\msun\ and
1.15\,\msun\ have been tested by \citet{sim2010a} (see also
\citet{shigeyama1992a}) and yield $^{56}$Ni masses in the range of
$[0.01\ldots 0.81]$\,\msun.  According to the model sequence of
\citet{sim2010a}, a standard normal SN~Ia with $\sim$\,0.6\,\msun\ of
$^{56}$Ni is expected to result from a detonation in a WD of about
1.1\,\msun.\ The observables predicted from these models roughly match
the data from normal SNe~Ia and their $B$-band light curves seem to
follow the width-luminosity relation \citep{sim2010a}. Thus,
detonations in WDs with masses well below the Chandrasekhar-limit hold
promise for explaining normal SNe~Ia. The question is how a detonation
in such an object can be triggered. Here we discuss one possibility
arising from a detonation in an accreted He shell on top of the
WD. Another possibility -- due to the merger of two WDs -- will be
presented in the next section.

The idea of double detonations in sub-Chandrasekhar mass WDs has been
discussed extensively in the 1990s by
\citet{woosley1994a,livne1995a,benz1997a,livne1997a, garcia1999a}. A
CO WD accretes helium from a companion star (either a helium star or a
helium WD). When the accreted He layer becomes sufficiently massive,
compressional heating is thought to lead to a detonation in the He
material (see, however, \citep{guillochon2010a} for an alternative
mechanism based on instabilities in the accretion process).  This
detonation sweeps around the CO core and burns the He to heavier
elements. At the same time a shock wave propagates into the core. This
shock may trigger a secondary detonation close to the interface
between the CO core and the He shell (``edge-lit detonation''), or
when reaching the center of the core. The secondary detonation
incinerates the entire WD and leads to its successful disruption in a
thermonuclear supernova. The question, however, is whether the event
would really look like a SN~Ia.  Although for sufficiently massive CO
cores enough $^{56}$Ni can be produced to power a normal SN~Ia,
problems arise from the burning products of the He shell. In the
models of the 1990s, a rather massive He shell -- about $[0.1\ldots
  0.2]$\,\msun\ -- was thought to be necessary to trigger a detonation
and to drive a sufficiently strong shock wave for initiating a
secondary detonation in the core. In such massive He shells, a
detonation produces a significant fraction of iron group elements
(including additional $^{56}$Ni). These affect the radiative transfer
and the predicted observables are at odds with the actual observations
\citep{hoeflich1996a, hoeflich1996b, nugent1997a,woosley2011b}.

\begin{figure}
  \centering
  \includegraphics[width=\linewidth]{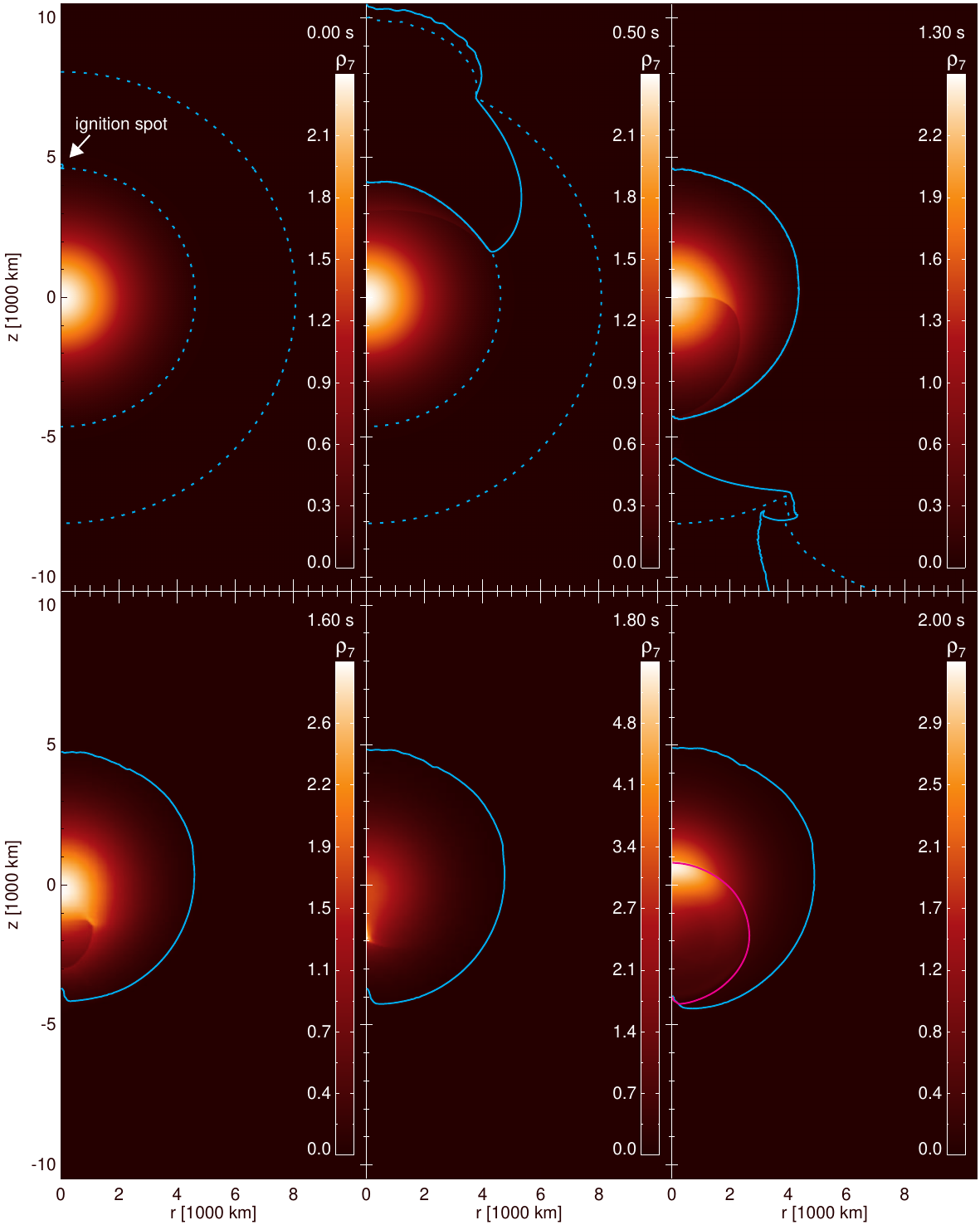}
  \caption{Hydrodynamic evolution of a sub-Chandrasekhar-mass
    double detonation for a helium shell of 0.084\,\msun\ on top
    of a 0.920\,\msun\ CO WD (from top left to bottom right; 
    model 2 of \citet{fink2010a}).  The density structure of the
    WD is colour coded in red ($\rho_7$ in units of $10^7$\,\gccm).  
    The dashed blue lines indicate the border of the helium shell. 
    The solid blue and magenta lines show the helium and CO 
    detonation flames, respectively.}
  \label{fig:hydro_subch}
\end{figure}

Recently, however, \citet{bildsten2007a} and \citet{shen2009a} pointed
out that in AM CVn systems rather low masses of accreted He on top of
a CO WD can develop dynamical burning, possibly in the
detonation mode. The work by \citet{fink2007a, fink2010a} demonstrated
that a core ignition is very robust due to spherical shock convergence
near the center of the WD which leads to a geometrical shock
amplification. Neither asymmetric ignition geometries
\citep{fink2007a} nor low He shell masses prevent a secondary core
detonation once the He shell successfully triggers a detonation
\citep{fink2010a}. The low He shell mass significantly reduces the
observationally disfavored effects of iron group elements in the outer
layers of the ejecta \citep{kromer2010a}. In addition to lowering the
total mass of the He shell, it also reduces the density at which He
detonates thus leading to predominantly incomplete
burning. Consequently, the outer layers of the ejecta in the models of
\citet{fink2010a} contain virtually no $^{56}$Ni and only low amounts
of other iron group elements. Another effect that distinguishes the
models of \citet{fink2010a} from many earlier calculations and also
from the recent models of \citet{woosley2011b} is the
multi-dimensional treatment of the He shell detonation. Sweeping
around the CO core, it propagates laterally and allows for
significant post-shock expansion -- an effect that is not captured in
spherically symmetric models. This adds to the less complete burning
observed in the simulations of \citet{fink2010a} (see also
\citep{townsley2012a}).  According to \citet{kromer2010a} the reduced
yields of heavy elements from the He detonation have significant
impact on the predicted observables. Although the colors are too red
to match the observations perfectly, the range of brightnesses and
rise and decline rates of normal SNe~Ia is covered by the models.

\citet{kromer2010a} also point out that the details of spectra and
the colors are very sensitive to the thermal and chemical conditions
in the detonating He shell.  In particular, they find that the degree 
of burning in the shell material (and thus its final composition)
can be affected by the initial composition of the He shell. Since 
the time-scale for $\alpha$-captures behind the detonation shock 
front is significantly shorter than that of triple-$\alpha$ reactions, 
a $^{12}$C admixture in the He shell due to previous hydrostatic 
burning or dredge-up of core material \citep{shen2009a} can limit 
the $\alpha$-chain before reaching nuclear statistical equilibrium. 
In an exploratory model \citet{kromer2010a} homogeneously polluted
a He shell with 34\% (by mass) of $^{12}$C (their model 3m) and showed
that such a model produces light curves and spectra that are in good 
agreement with those of normal SNe~Ia. In particular, this model
is no longer too red at maximum light. However, one caveat remains:
the model still produces a non-negligible amount of Ti in the outer
layers leading to the formation of a \ions{Ti}{ii} absorption trough
between 4000\,\AA\ and 4400\,\AA\ which is not observed in normal
SNe~Ia (see also middle panel of Fig.~\ref{fig:specplot_normals})
but in subluminous 91bg-like supernovae only 
(Sect.~\ref{sec:observed_diversity}).
Whether these differences can be resolved as well, remains to be seen
in future studies that more fully explore the influence of the initial 
composition of the helium shell and different ignition geometries.
Also the strong sensitivity of the radiative transfer to tiny amounts
of particular elements requires a better description of nuclear 
reaction rates and continued study of the radiative transfer 
processes (and atomic data) in order to quantify more fully the 
systematic uncertainties which arise due to the complexity of spectrum 
formation in supernovae. 

\subsubsection{Violent mergers}
\label{sec:violent_mergers}

Another external trigger to ignite a sub-Chandrasekhar-mass WD is the
violent merger of two CO WDs. Although the total mass of the merging
system usually exceeds the Chandrasekhar mass, both components are
below this mass limit. The so-called \emph{violent merger model}
\citep{pakmor2010a} starts with massive WDs ($M\gtrsim 0.9$\,\msun\,)
with a mass ratio close to unity. This scenario results from a subset
of double-degenerate progenitor models.  Other configurations of
merging WDs may avoid thermonuclear explosions
(e.g., \citep{saio1998a}) and instead lead to the formation of a
neutron star by gravitational collapse \citep{saio1985a}.

\begin{figure}
  \centering
  \includegraphics[width=\linewidth]{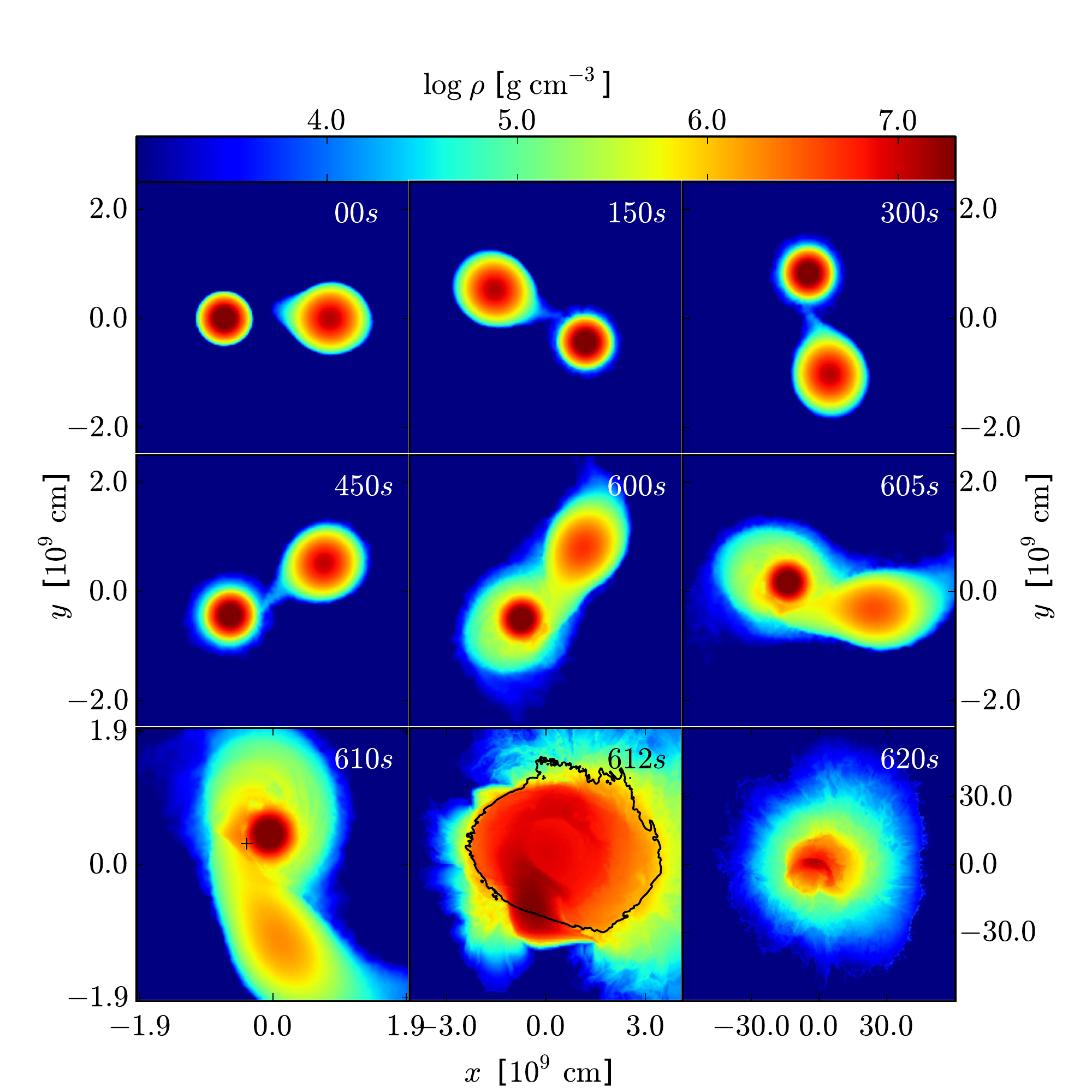}
  \caption{Hydrodynamic evolution and subsequent thermonuclear explosion
    of a merger of a pair of 1.1 and 0.9 \msun\ WDs \citep{pakmor2012a}. 
    Initially the WDs orbit each other with a period of $\sim$\,35\,s.
    After a few orbits the secondary is tidally disrupted and collides
    with the more massive primary reaching densities and temperatures
    sufficient to ignite a detonation at 610\,s (black cross). At 612\,s
    the detonation front (black line) has burned almost the complete
    object. Color-coded is the logarithm of the density. Note that the 
    last two panels have a different color scale ranging from $10^{-4}\,
    \mathrm{g\ cm^{-3}}$ to $10^{6}\,\mathrm{g\ cm^{-3}}$ and 
    $10^{4}\,\mathrm{g\ cm^{-3}}$, respectively.}
  \label{fig:evo_merger}
\end{figure}

For mass ratios close to unity, however, the mergers proceed
dynamically and can be followed in hydrodynamic
simulations. \citet{pakmor2012a} studied the merger of a
1.1\,\msun\ ``primary'' WD with a 0.9\,\msun\ ``secondary'' WD.  The
inspiral and merger, as followed with the SPH code \textsc{Gadget}
\citep{springel2005a} in its modification for stellar astrophysical
problems \citep{pakmor2012b} is shown in Fig.~\ref{fig:evo_merger}.
In the last few orbits before the actual merger, tidal interaction
strongly deforms the secondary and it finally plunges into the primary
WD (snapshots for $t > 600\,\mathrm{s}$ in Fig.~\ref{fig:evo_merger}).
This violent merger leads to the formation of a hot spot where the two
masses collide (marked by a black $+$ in the snapshot for $t =
610\,\mathrm{s}$). Here, thermodynamic conditions are suitable for
triggering a detonation.

After mapping into our grid-based supernova explosion code, the
detonation (indicated by a black contour in Fig.~\ref{fig:evo_merger})
is followed with the level-set technique (see
Sect.~\ref{sec:models_hydro}). It incinerates the merged object almost
completely. An important point to notice is that the primary WD is
nearly unaffected by the merger. Therefore the burning takes place at
the low densities typically found in sub-Chandrasekhar-mass WDs. Only
the primary possesses material at sufficiently high densities to
synthesize iron group elements while the secondary mostly burns to
oxygen. With the mass of the primary chosen to 1.1\,\msun,\ a moderate
$^{56}$Ni mass production is expected according to \cite{sim2010a} and
indeed 0.64\,\msun\ of $^{56}$Ni are found in the presented
simulation.  Significant amounts of the total 2\,\msun\ of material
involved in this merger burn to intermediate-mass elements
(0.6\,\msun)\ and the ejecta contain 0.47\,\msun\ of oxygen. Only
0.09\,\msun\ of carbon remain in the ejecta.

Thus, despite the large total mass of the exploding object, the
angle-averaged and line-of-sight dependent light curves of this merger
compare very favorably to that of normal SNe~Ia.  With a peak
brightness of $-19.6$, $-19.0$, and $-19.2$ in the $U$, $B$, and $V$
bands, respectively, and a $B$-band light curve decline rate of
$\Delta m_{15}(B)=0.95$, the model predictions are well in the range
of those observed for normal SNe~Ia \citep{hicken2009b}. Moreover, the
spectral evolution of the model (see Fig.~\ref{fig:specplot_normals})
reproduces the overall spectral shape and the velocity-shifts of most
of the line features remarkably well and shows most of the
characteristic features of SNe~Ia, particularly the defining Si\,{\sc
  ii} doublet at $\lambda\lambda$6347,6371 but also the weaker
Si\,{\sc ii} features at $\lambda\lambda$5958,5979 and $\lambda\lambda
$4128,4131. Other prominent features are the Ca\,{\sc ii} H and K
absorptions, the Mg\,{\sc ii} triplet at $\lambda$4481, the S\,{\sc
  ii} W-feature at $\sim$5400\,\AA\ and, in the red tail of the
spectrum, the O\,{\sc i} triplet $\lambda\lambda$7772,7774,7775 and
the Ca\,{\sc ii} NIR triplet. However, with a $B$-band rise time of
20.8\,d the pre-maximum light curve evolution of this merger model is
relatively long compared to that of normal SNe~Ia.
\citet{hayden2010a}, for example, find an average $B$-band rise time
of 17.4\,d in the SDSS-II SN sample (but see also
\citealt{conley2006a} who find a value of 19.58\,d for a low red-shift
sample from the SNLS). This could indicate that the total ejecta mass
in this model is somewhat too large. Future studies exploring the
parameter space of violent WD mergers in more detail will show if 
this explanation is right.

\subsubsection{Critical assessment}
\label{sec:assess}

Although the models presented above produce successful explosions that
overall compare favorably to the observations of normal SNe~Ia, there
remain uncertainties in the modeling of the explosion physics. In all
models this refers to the initiation of the burning and the formation
of detonations. This is not too surprising as these
processes work on scales that cannot be resolved in our
multi-dimensional supernova simulations. Moreover, the ignition of
deflagrations and detonations are complex physical phenomena and their
microphysics not
completely understood, even for terrestrial combustion. Despite our
attempt to model the explosion physics as parameter-free as possible, 
we are thus left with the following critical points in the three 
classes of models:

\begin{itemize}
\item \emph{Delayed detonations in Chandrasekhar-mass WDs} hinge on
  the possibility of deflagration-to-detonation transitions to occur 
  in WD combustion. Although some recent studies
  (e.g., \cite{roepke2007d,woosley2007a,woosley2009a,poludnenko2011a})
  indicate that this may indeed be the case, it is difficult to
  definitely decide on its realization in SNe~Ia. The other major
  uncertainty in this model is the way the deflagration ignites. A strongly
  asymmetric ignition leads to extremely bright events in the context of
  delayed detonations. Reaching the low-luminosity end of the normal
  SNe~Ia requires to limit the $^{56}$Ni production to the yield of the
  strongest pure deflagrations. These result from quasi-isotropic or
  central ignitions of the WD -- a scenario that is currently not
  favored by ignition simulations \citep{zingale2009a} but may arise due
  to slight rotation in the ignition phase \citep{kuhlen2006a}.
\item \emph{Double detonations in sub-Chandrasekhar mass WDs} require
  the initiations of two detonations. While the secondary detonation
  in the CO core seems to be virtually unavoidable
  \citep{fink2007a,fink2010a}, the initial detonation in the He shell
  is not established beyond doubt -- in particular for the case of
  low He-shell masses.
\item \emph{Violent mergers of two WDs} rely on the triggering of a
  detonation at the encounter of the two stars. Although the
  simulations of \citet{pakmor2010a, pakmor2011b,pakmor2012a} indicate
  that this is possible, the mechanism still awaits a firm proof. 
\end{itemize}

A better understanding of the microphysics of thermonuclear combustion
is thus required to overcome these uncertainties and to assess the
models purely from the plausibility of their explosion mechanism.
This is a challenging task and a convincing result is not expected in
the short term. There are, however, alternative ways to judge the
potential of different explosion scenarios to account for the majority
of normal SNe~Ia. 

One possibility, which we followed here, is to perform supernova
simulations under the assumption that the uncertain mechanisms in the
modeling proceed in a favorable way and to compare the outcome with
observations. Our results indicate that all of the models considered
here are able to cover the range of explosion energies and
brightnesses of normal SNe~Ia and to first order reproduce their light
curves and spectra relatively well. The comparisons of our synthetic
spectral time series and SN~2005cf in Fig.~\ref{fig:specplot_normals}
demonstrate this success, but also show that in detail there are
shortcomings in each of the models as discussed in the previous
sections. In the Chandrasekhar-mass delayed-detonation and
sub-Chandrasekhar-mass double-detonation models the blue-shift of the
characteristic \ions{Si}{ii} is too large compared to SN~2005cf,
indicating slightly too high ejecta velocities. This potentially can
be cured by more realistic progenitor models with carbon-depleted
cores \citep{roepke2012a}. Moreover, the models are too red compared
to the observations. This is most pronounced for the
delayed-detonation Chandrasekhar-mass model but also found in the
double-detonation sub-Chandrasekhar-mass and the violent merger model
to some degree.

Although the involved masses and the explosion physics of the models
shown in Fig.~\ref{fig:specplot_normals} differ significantly, at the
current precision of the models it is difficult to distinguish them
by means of maximum-light optical spectra only. This degeneracy prevents
favoring one model over others. This could imply that all channels
contribute to normal SNe~Ia (possibly with different realization
frequencies) or other ways of discriminating them have to be
found. Promising for this task seem observations in the ultraviolet
(e.g., \citep{lentz2000a,foley2012a}) and in the near-infrared bands
(e.g., \cite{marion2003a}) but also late time observations (e.g.,
\citep{kozma2005a, roepke2012a}), spectropolarimetry
(e.g.,\citep{kasen2003a, wang2008a}) or gamma-ray observables (e.g.,
\citep{gomez-gomar1998a, sim2008a, maeda2012b}). For these, either
theoretical models have yet to be developed for modern
multi-dimensional SN~Ia simulations, or data has to be acquired.

Other possibilities to discriminate different explosion scenarios 
are the search for signatures of the progenitor 
system in nearby SNe~Ia (see, e.g., \citep{nugent2011a, bloom2012a, 
li2011a, liu2012b, brown2012a, chomiuk2012a, horesh2012a} 
for observational constraints and, e.g., \citep{marietta2000a, 
pakmor2008a, pan2012a, liu2012a} for theoretical predictions) or in 
supernova remnants (e.g., \citep{ruiz-lapuente2004a,kerzendorf2009a,
kerzendorf2012a,schaefer2012a}).

Finally, any model scenario that is claimed to account for a large
fraction of SNe~Ia must be able to explain observational trends like
e.g.\ the observed delay-time and brightness distribution of SNe~Ia.
By combining the synthetic observables from our explosion models with
studies of the realization frequency of the supposed progenitor
systems, as discussed in Sect.~\ref{sec:progenitors}, we can thus put
additional constraints on the different explosion scenarios.

Recently, we have used this approach to investigate the prospect of
the violent merger scenario in more detail.  As was noted in
Sect.~\ref{sec:violent_mergers}, the secondary WD in a violent merger
-- while consumed in the explosion -- does not contribute to
synthesizing $^{56}$Ni.  Thus, it is the primary WD (more specifically
its mass) that simply determines the peak luminosity of a SN Ia in the
violent merger model.

For a realistic estimate of primary WD masses in would-be merging WD
pairs, we took the distribution of primary WD masses of all merging CO
WDs from the binary evolution population synthesis calculations of
\citet{ruiter2011a} (their standard model).  Using this mass
distribution, a relationship between the (primary, sub-Chandrasekhar
mass) WD and its corresponding SN bolometric peak brightness ($m_{\rm
  WD}-M_{\rm bol}$) was derived using the technique as described in
\citet{sim2010a}.

One critical question is the realization of a WD merger itself:
e.g.~what is the critical mass ratio for which mass transfer will be
dynamically unstable (and lead to a merger) when the larger WD fills
its Roche-lobe?  The answer to this question is not straightforward,
and much uncertainty exists in the modeling of mass transfer in close
binaries \citep{marsh2004a}.  A trend that was found in the merger
simulations of \citet{pakmor2012a} (see also \citep{han1999a}) 
is that the critical mass ratio
$q_{\rm c}$ likely decreases with larger primary masses.  We
constructed a relationship (\citep{ruiter2012a}, equation 1) that
follows this trend to evaluate whether a given double WD could produce
a merger that is sufficiently violent.  Additionally, for the violent
WD mergers we limited the primary mass to be above $0.8$ \msun\, since
primaries less massive than this are expected to barely produce even
0.01 \msun\ of $^{56}$Ni (\citep{sim2010a}, table 1).

\begin{figure}
  \centering
  \includegraphics[width=\linewidth]{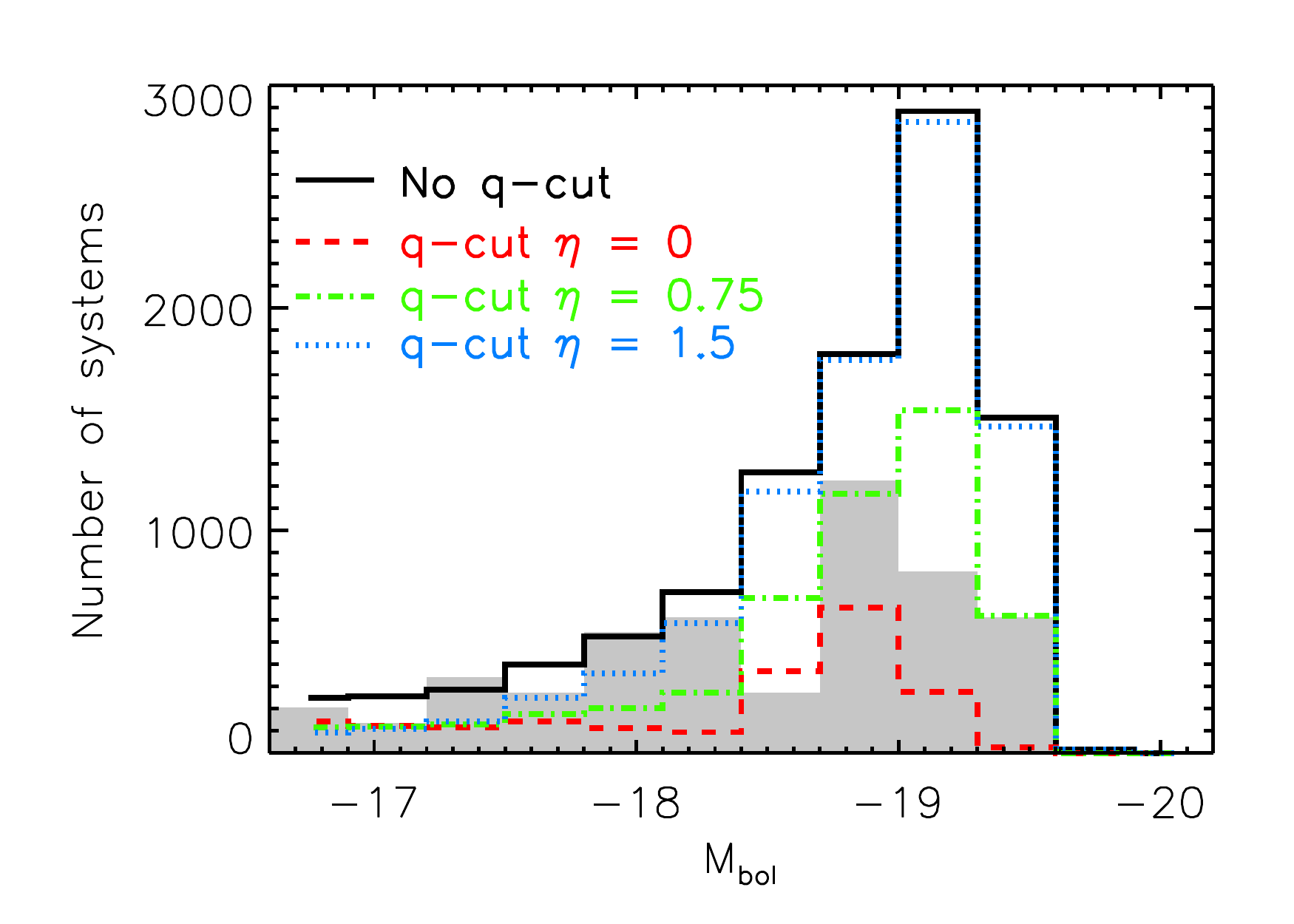}
  \caption{Brightness distribution of violent WD mergers
    \citep[see][for details]{ruiter2012a}.  Black solid histogram
    shows {\em all} CO WD mergers from population synthesis, while
    coloured histogram lines show the brightness distributions when
    more stringent mass ratio constraints are assumed.  Grey scale
    shows the observational peak brightness distribution of $74$ SNe
    Ia from the volume-limited sample of 
    \citet{li2011a}; observations are scaled up to enable comparison
    with the distribution shapes from our models. }
  \label{fig:Mp-Mbol}
\end{figure}

Fig.~\ref{fig:Mp-Mbol} shows four model (bolometric) peak brightness 
distributions for a range of $q_{\rm c}$-cuts.  In grey scale we
over plot the (scaled-up) observational luminosity distribution of
SNe Ia from \citet{li2011a}.  Regardless of the assumed $q_{\rm c}$-cut, our
theoretical brightness distributions do a fairly good job in covering
the range and matching the shape of the observed SN Ia brightness 
distribution.  Such good agreement indicates that merging WDs 
which explode via the violent merger mechanism could be dominant 
SN Ia progenitors, driving the shape of the 
underlying brightness distribution.  

\begin{figure}
  \centering
  \includegraphics[width=10cm]{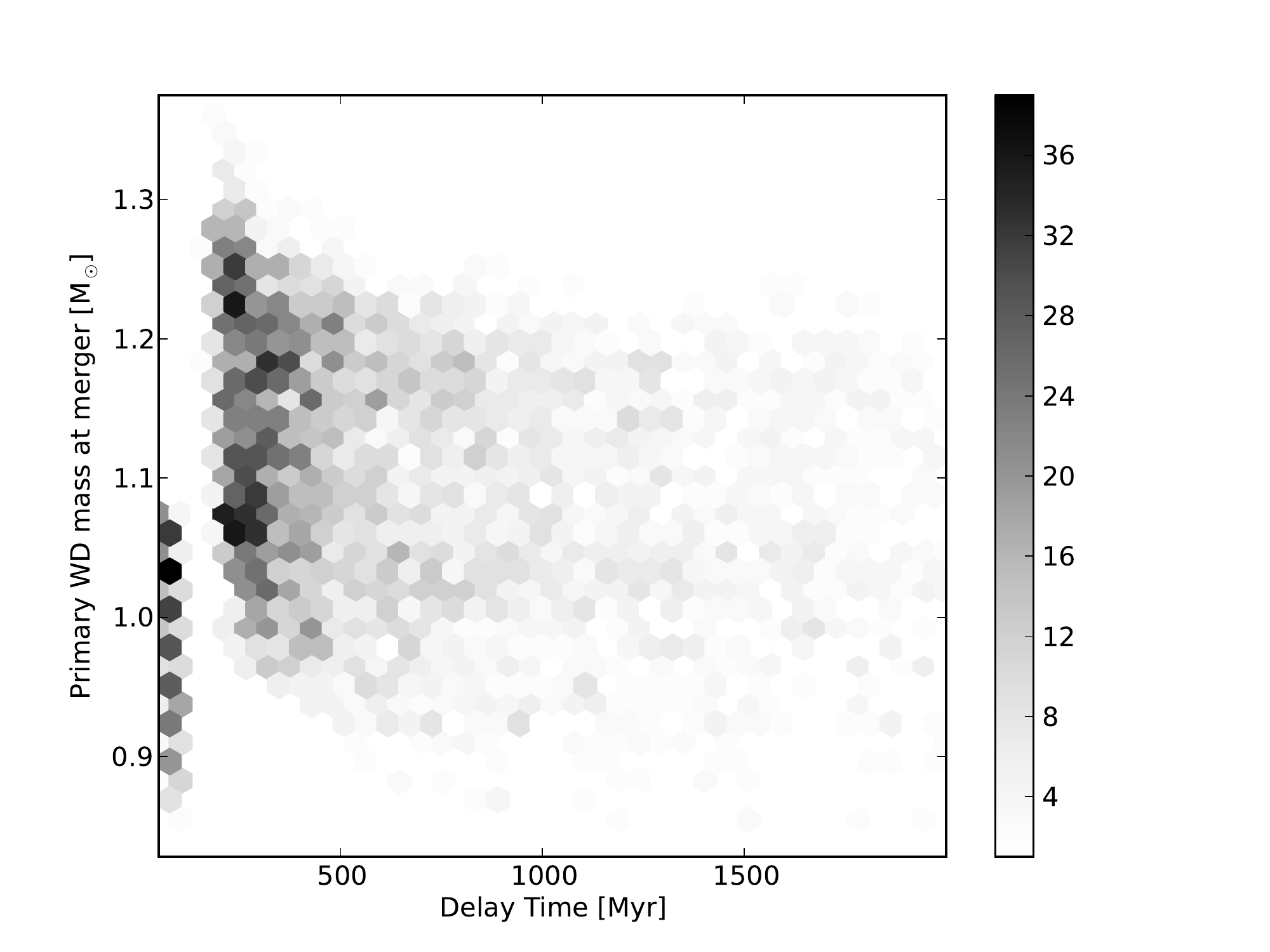}
  \caption{Delay time distribution for primary WD masses at time of
    merger for one of our violent merger model populations (cf.~blue
    histogram in Fig.~\ref{fig:Mp-Mbol}).  We show delay times
    only from $0 - 2000$ Myr so that the characteristic primary masses
    at prompt delay times are clearly visible.  The most massive
    primaries tend to merge at delay times $< 400$ Myr, however there also
    exists a distinct population of `ultra-prompt' mergers with
    less-massive primaries with delay times $<100$ Myr (see text).
}
  \label{fig:Mp-DTD}
\end{figure}

In Fig.~\ref{fig:Mp-DTD} we show the primary mass in the same model
for violent WD mergers as a function of delay time.  The darkest
hexagons represent the regions of highest density for a given cell.
It is clear from the plot that mergers hosting the most massive
primaries ($\gtrsim 1.3$ \msun) tend to merge at prompt ($< 500$ Myr)
delay times.  
(We note that in single star evolution, CO~WDs would not achieve such 
high masses and a WD of mass ${\sim}1.3$ \msun\, would be composed 
mostly of oxygen and neon.  However, in binary evolution, such 
masses are allowed for CO~WDs, in particular if the CO~WD accretes 
mass after it has formed (see e.g. \citep{ruiter2012a}, fig.\ 2)). 
Since in the violent merger model the SN Ia luminosity
is determined by the primary WD mass, this means that we would expect
the brightest SNe to be found amongst very young stellar populations;
a trend which is confounded by observations \citep{brandt2010a}.  One
aspect of the binary evolution model which remains to be confirmed, is
whether or not primary WDs are able to efficiently accrete on the
order of $0.2$ \msun\ from a slightly-evolved helium star companion.
Such a mass transfer phase was found to be critical in producing a
large number of primary WD masses that yield peak explosion
brightnesses around $-19$ mag (see \citep{ruiter2012a}).
Additionally, we note that there exists another population of violent
mergers with very short ($<100$ Myr; ultra-prompt) delay times.  These
systems undergo two common envelope phases, whereby the secondary star
loses its envelope twice.

\subsection{Peculiar SNe}
\label{sec:peculiars}

As discussed in Sect.~\ref{sec:observed_diversity}, as of today 
several peculiar sub-classes of SNe~Ia have been found in addition
to the bulk of spectroscopically normal SNe~Ia  
which follow the Phillips relation.  Here, we
discuss possible explosion models for a few of those peculiar 
sub-classes.

\subsubsection{1991bg-like SNe}

1991bg-like SNe are subluminous with respect to the Phillips relation
and peak at about -17\,mag, indicating that only a rather low 
\nuc{56}{Ni} mass of about 0.1\,\msun\ was synthesized during the
explosion.  Moreover, a spectral analysis of SN~2005bl, a well-observed
proto-typical 1991bg-like SN, has shown that both iron group elements 
and silicon are present over a wide range of radii extending down to
very low expansion velocities.  This indicates the presence of 
incomplete Si burning over a wide velocity range in these explosions
as it may occur in detonations at low densities.

\begin{figure*}
  \centering
  \includegraphics[width=\linewidth]{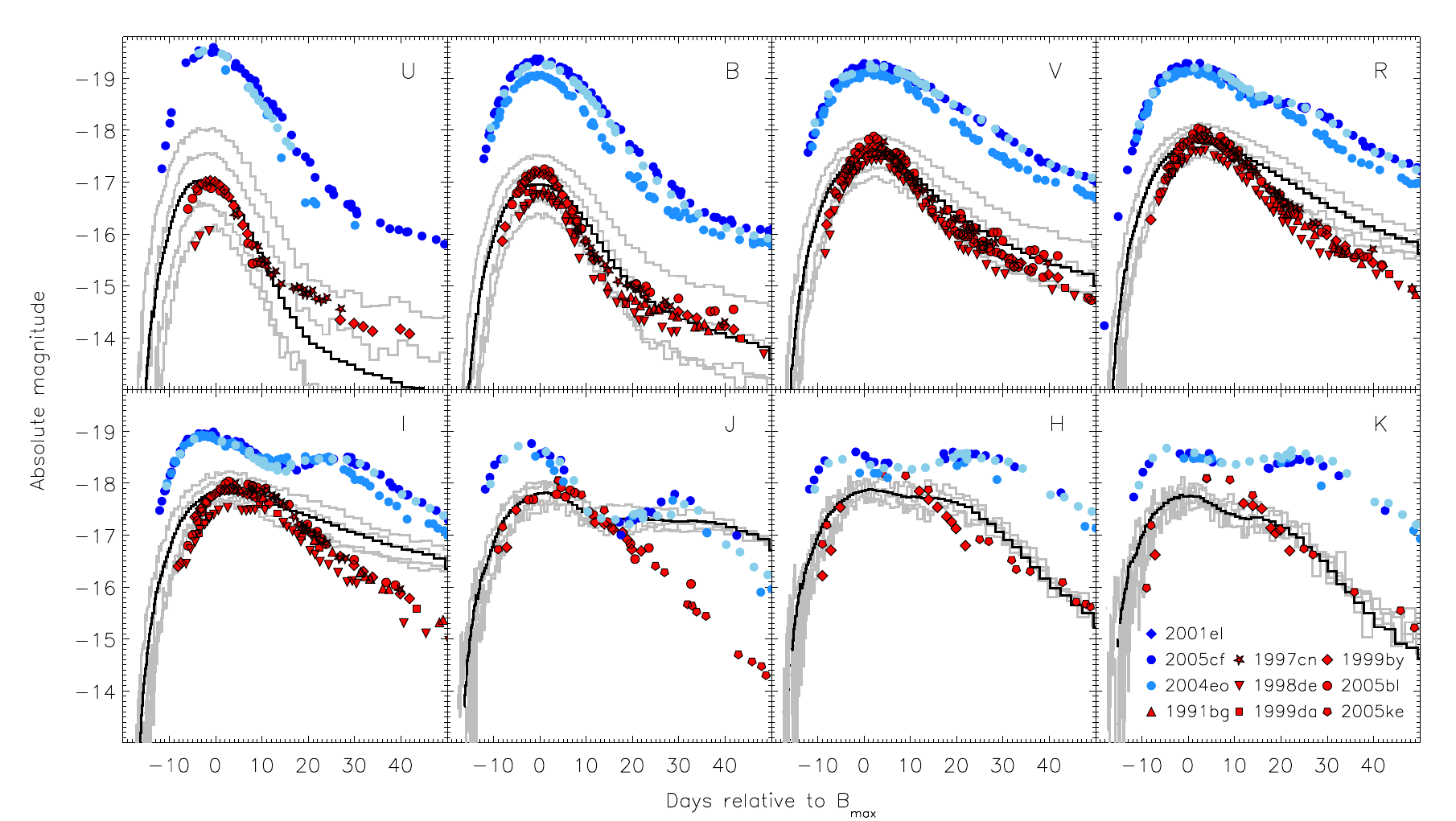}
  \caption{Angle-averaged synthetic light curves of a merger of two 
    0.89\,\msun\ WDs (black).  To indicate the spread due to different
    viewing angles, the gray lines show light curves along four 
    different lines-of-sight. These have been selected from 100 
    equally sized solid-angle bins such that they represent the full
    range of the scatter. For comparison observed photometry for 
    normal (blueish colours, \citet{krisciunas2003a, pastorello2007a, 
    pastorello2007b}) and sub-luminous 1991bg-like SNe (red,
    \citet{taubenberger2008a} and references therein) is shown.}
  \label{fig:specplot_1991bg}
\end{figure*}

In the violent merger scenario (see also 
Sect.~\ref{sec:violent_mergers} \citep{pakmor2010a}) such a burning is
possible for a primary WD with a sufficiently shallow density profile.
Following the inspiral of a pair of 0.89\,\msun\ WDs with the SPH code
{\sc gadget} and using our full modeling pipeline \citet{pakmor2010a} 
have shown that such a configuration produces about the right amount
of \nuc{56}{Ni} although a total mass of $1.8\,$\msun\ is involved in 
the merger.  Moreover, their simulation can reproduce the observed 
spectra and light curves of 1991bg-like SNe (see 
Figure~\ref{fig:specplot_1991bg}) and accounts for most of their 
peculiar features.

\citet{pakmor2011b} find that WD binaries with a primary mass of
$M_1\sim$\,0.9\,\msun\ and mass ratios $q=M_2/M_1>0.8$ evolve
similarly, thus confirming a robust ignition of $\sim$\,0.9\,\msun\
violent WD mergers which makes them promising candidates for 1991bg-like
SNe given the good agreement of synthetic observables in our pilot
study.  Arguing that primary WDs with a lower mass will not detonate
due to their lower densities and using population synthesis 
calculations of \citet{ruiter2009a}, \citet{pakmor2010a} also 
estimated the rate of binary mergers that met the necessary 
criteria to satisfy their model.
It was found that such mergers may contribute on the order of $2 -
11$ \% to the total SN~Ia rate, which is not too far off from the
observationally derived rate for 1991bg-likes of 15\% \citep{li2011a}. 
Moreover, if the WD binaries undergo 
only one common envelope phase and/or begin their evolution on the 
ZAMS with wide orbital separations their model 
also prefers old ($>1$\,Gyr) stellar populations as indicated by 
observations of 1991bg-like SNe.

\subsubsection{2002cx-like SNe}
\label{sec:2002cx-likes}

One of the most peculiar sub-classes of SNe~Ia are explosions similar
to SN~2002cx \citep{li2003a}.  Those events are under-luminous with
respect to the Phillips relation and their NIR light curves do not
show secondary maxima. Moreover, their spectra are characterized by
very low expansion velocities compared to normal SNe~Ia and show signs
of strongly mixed ejecta.  While explosion models involving a
detonation are not able to explain such an ejecta structure
(e.g., \cite{sim2010a, seitenzahl2012a}), turbulent deflagrations in
Chandrasekhar-mass WDs naturally predict such a strong mixing and low
kinetic energies \citep{gamezo2004a, roepke2005b}.

\begin{figure*}
  \centering
  \includegraphics[width=0.8\linewidth]{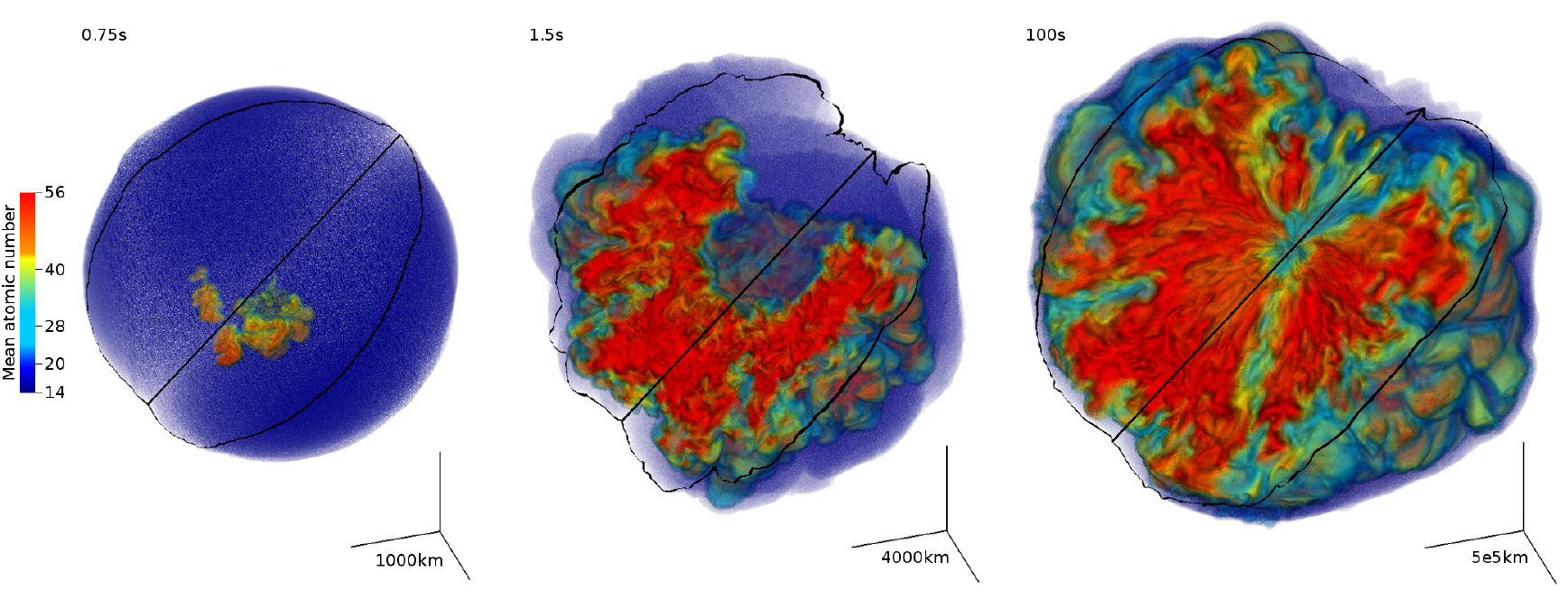}
  \caption{Hydrodynamic evolution of an asymmetrically ignited
    deflagration in a Chandrasekhar-mass WD. Particularly we 
    show model N5def of Fink et al. (in preparation) which 
    ejects about 0.37\,\msun\ of which $\sim$\,0.16\,\msun\ 
    are \nuc{56}{Ni}.  Shown is a volume rendering
    of the mean atomic number (colour bar) from which we carved 
    out a wedge to allow a view into the core.  At 0.75\,s after 
    the explosion a one-sided deflagration plume rises towards 
    the WD surface which fragments due to buoyancy instabilities.  
    (ii) At 1.5\,s the expansion of the WD quenches the burning 
    and the explosion ashes wrap around the unburned core.  
    (iii) Finally, at 100\,s the unburned core is completely 
    engulfed by the explosion ashes which are accelerated to 
    escape velocity.
    }
  \label{fig:hydro_failed_deflag}
\end{figure*}

Given our ignorance of the exact ignition configuration of 
Chandrasekhar-mass WDs (see Sect.~\ref{sec:normals_from_MCh}),
we have recently performed a systematic study of 3D full-star 
explosion simulations of pure deflagrations in Chandrasekhar-mass 
WDs (Fink et al., in preparation) for different ignition setups.
Depending on the strength of the ignition which is parametrized
by a varying number of ignition sparks to seed unstable burning 
modes we obtain \nuc{56}{Ni} masses between 0.035 and 0.38\,\msun.\
Moreover, we find that only strong ignitions release enough energy
during the burning to unbind the progenitor WD completely. 
Asymmetric, weak ignition setups, in contrast, lead to a one sided
deflagration plume which fragments due to Rayleigh-Taylor and 
Kelvin-Helmholtz instabilities and finally wraps around the still
unburned WD core when it comes close to the surface 
(Figure~\ref{fig:hydro_failed_deflag}; see also \citep{plewa2004a, 
roepke2007a}). However, even deflagrations which fail to unbind the
complete WD accelerate parts of their explosion ashes to escape
velocity and eject this material into their surroundings.

\begin{figure}
  \centering
  \includegraphics[width=0.66\linewidth]{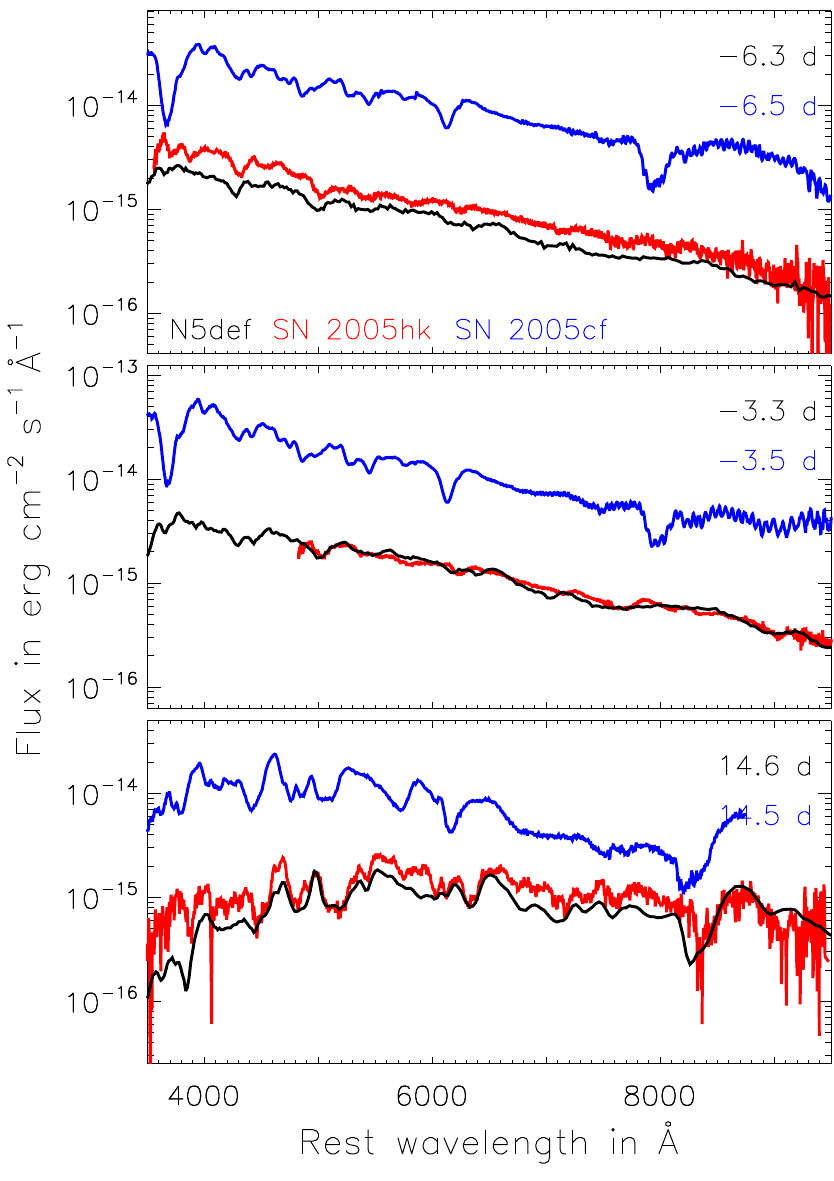}
  \caption{Synthetic spectra of the asymmetrically ignited deflagration 
    model N5def which leaves behind a bound remnant. The spectral 
    evolution is remarkably similar to SN~2005hk \citep{phillips2007a}, 
    a proto-typical 2002cx-like SN. For comparison we show also spectra
    of SN~2005cf \citep{garavini2007a} as an example for a normal SN~Ia.}
  \label{fig:specplot_failed_deflag}
\end{figure}

Using a million Lagrangian tracer particles, we determined the
detailed chemical composition of our simulations from a
post-processing calculation with our 384-isotopes nuclear network
\citep{travaglio2004a, seitenzahl2010a} and mapped the resulting
ejecta structure into our radiative transfer code {\sc artis}
\citep{kromer2009a, sim2007b}.  While the obtained synthetic
observables for strong deflagrations, which completely unbind the
progenitor WD, do not match the display of observed SNe, deflagrations
which leave behind a bound remnant closely resemble the observed
properties of 2002cx like SNe
(Figure~\ref{fig:specplot_failed_deflag}, for details see
\citet{kromer2012a} and \citet{jordan2012b}).

\subsubsection{Superluminous or ``super-Chandra'' SNe}

Recently observations revealed a new class of superluminous SNe~Ia 
(e.g.,\citep{howell2006a,taubenberger2011a}) with total ejecta 
masses significantly larger than the canonical Chandrasekhar mass 
of 1.4\,\msun.\  SN~2009dc, which is one of those objects, requires 
even a \nuc{56}{Ni} mass larger than the Chandrasekhar mass if its 
peak luminosity was solely powered by radioactive decay of 
\nuc{56}{Ni} and its daughter nuclei \citep{taubenberger2011a, 
kamiya2012a}.  One model which was proposed for these peculiar
objects by \citet{howell2006a} is that of exploding rapidly 
rotating WDs which stay stable well above 1.4\,\msun due to 
centrifugal forces \citep{yoon2005b}.  

\begin{figure}
  \centering
  \includegraphics[width=\linewidth]{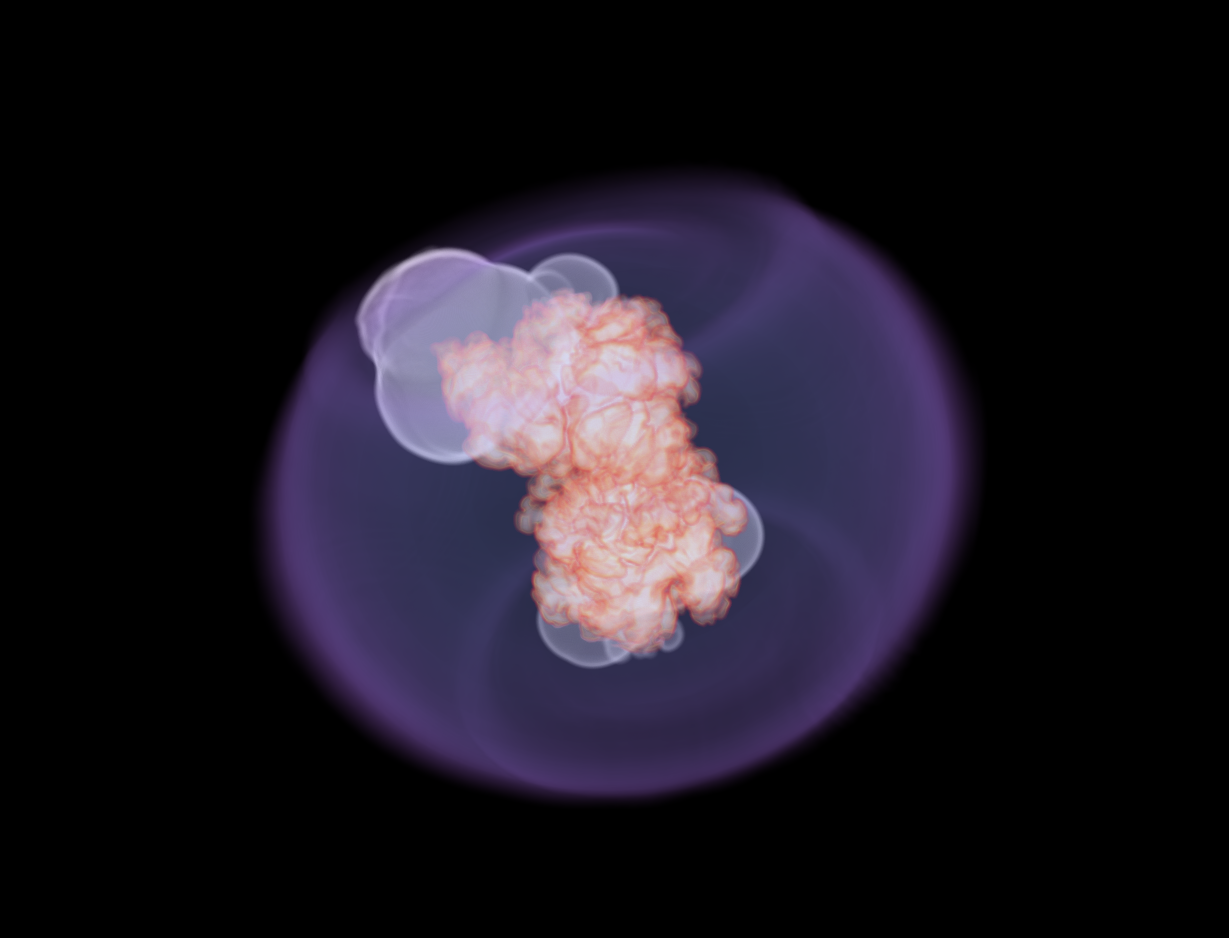}
  \caption{Delayed detonation in a differentially rotating WD 
    of 2\,\msun.\  Shown is a snapshot of the flame evolution 
    at 0.9\,s after the explosion.  It is clearly visible that
    the propagation of the wrinkled deflagration flame (reddish
    surface) is inhibited in lateral directions by the rotation
    and propagates predominantly along the rotation axis.  
    At several location a deflagration-to-detonation transition
    occurred and detonation flames started to spread (whitish 
    surface).  The donut structure of the differentially rotating 
    WD is indicated by the blueish volume rendering of its density.}
  \label{fig:hydro_rotWD}
\end{figure}

Several authors have studied prompt detonations \citep{steinmetz1992a,
pfannes2010b} and turbulent deflagrations \citep{pfannes2010a} in such
differentially rotating WDs.  Here we report on a \emph{delayed 
detonation in a differentially rotating WD} of 2\,\msun\
(see Figure~\ref{fig:hydro_rotWD}; Fink et al., in preparation).  
Compared to a delayed detonation in a non-rotating WD, the initial 
deflagration propagates preferentially along the rotation axis since 
angular momentum conservation and weaker gradients in the effective 
potential inhibit the growth of flame instabilities in lateral 
directions.  A similar effect was already found by \citet{pfannes2010a} 
for pure deflagrations in such an object.  As a consequence not much 
energy is released during the deflagration phase.  Therefore, the WD 
does not expand strongly before the deflagration-to-detonation 
transition leaving a large amount of fuel at high densities which
the ensuing detonation efficiently burns to nuclear statistic 
equilibrium.

\begin{figure}
  \centering
  \includegraphics[width=\linewidth]{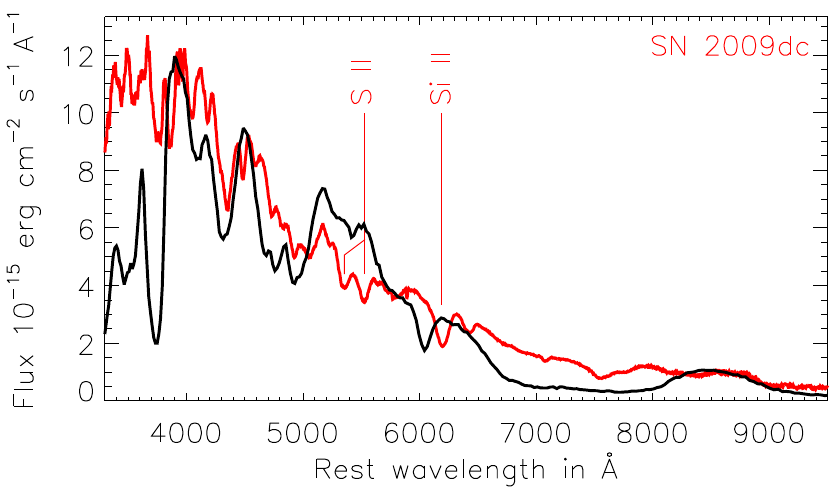}
  \caption{Synthetic maximum light spectrum of a delayed detonation
    in a differentially rotating WD of 2\,\msun.\  The characteristic
    absorption features of Si and S are blue-shifted with respect to
    the observed spectrum of SN~2009dc \citep{taubenberger2011a}, a 
    proto-typical super-luminous SN~Ia.  The binding energy of the 
    WD is not large enough to compensate for the huge energy release 
    due to nuclear burning leading to a too high kinetic energy.}
  \label{fig:specplot_rotWD}
\end{figure}

Yielding a total \nuc{56}{Ni} mass of 1.45\,\msun\ our simulation
gives rise to a bright explosion which in principle qualifies the
model as an explanation of super-luminous SNe~Ia.  However, the
observationally derived ejecta structure of those objects 
\citep{hachinger2012a} does not match our explosion.  This is
also reflected by the synthetic observables from our model which
do not match the observed spectra of super-luminous SNe~Ia
(see Figure~\ref{fig:specplot_rotWD}).  In particular absorption 
features of intermediate-mass elements, such as Si and S, are 
significantly blue-shifted with respect to the observed spectra, 
thus indicating that these elements are located at too large 
velocities in our model.  Moreover, we cannot reproduce the 
characteristic C features of super-luminous SNe~Ia in our model
since the detonation burns almost all the fuel.

As an alternative explanation for super-luminous SNe~Ia 
\citet{howell2006a} proposed the merger of two massive CO 
WDs.  However, at least in the violent WD merger scenario 
\citep{pakmor2010a}, this seems to be unlikely.  In this model the 
produced \nuc{56}{Ni} mass depends only on the mass of the primary WD.  
Since exploding CO WDs in the violent merger model usually 
have masses well below 1.3\,\msun\ \citep{ruiter2012a}, this 
essentially limits the achievable \nuc{56}{Ni} mass in a violent 
merger to $\sim$\,1\,\msun.\  For an alternative explanation of 
super-luminous SNe~Ia in an interaction scenario see 
\citet{hachinger2012a}.

\section{Summary and conclusions}

In this article we have reviewed some of the recent work on SNe Ia
done by the MPA-Garching group. Most of this work was motivated by the
fact that this class of stellar explosions is not as homogeneous as it
appeared to be in the past. In fact, new detailed observations of many 
nearby events as well as results from recent supernova surveys seem to 
indicate that there is not a single progenitor channel but that several
distinctively different channels are more likely.

Therefore we have started a new effort to simulate not only 
single-degenerate Chandrasekhar-mass explosions but also
sub-Chandrasekhar mass models and (violent) double-degenerate mergers,
to compute synthetic light curves and spectra from these models, and
to compare their predictions with data. Moreover, we have made an
attempt to compute rates and delay times of the different progenitor
classes from binary-population synthesis models. The main results of
this program were presented in the previous sections.

We have demonstrated that some of the models are able to reproduce
light curves and spectra of 'normal' SNe Ia amazingly well, given the
fact, that these models have almost no tunable (non-physical)
parameters. The agreement is not perfect but, given the uncertainties
still present in the models (initial conditions, combustion physics,
radiative transfer, ...) this is not so surprising. Also, it may be
better not to compare an individual supernova with a particular
realization of a special group of models but try to reproduce
'generic' features of a full class of objects instead. This will
become possible in the future once extended grids of models have been
computed.

The bad news is that rather different explosion models reproduce the
data equally well (or not so well). This 'degeneracy' can be
understood from the fact that mainly the abundances and distribution
of radioactive $^{56}$Ni and intermediate-mass nuclei determine the
observed properties of thermonuclear supernovae, and they are not too
different for delayed-detonation Chandrasekhar-mass models, for
sub-Chandrasekhar-mass explosions, or for violent mergers. These
classes of models differ mainly in their ejecta masses and the amount
of unburnt carbon and oxygen. But since the opacity of C and O is low
this has, in general, little effect on the light curves and spectra.
In the future, strong arguments in favor of one or the other
progenitor channel may come from constraints on the rates and delay
times. Our population synthesis models support the double-degenerate 
scenario, but this is still controversial. Other constraints come from
direct observations, such as the presence or absence of circumstellar gas, 
the non-detection of the progenitor star or its companion, and so on.
As it stands, the results of such studies are conflicting, but could
best be explained by more than one progenitor channel.

As far as some of the peculiar SNe~Ia are concerned models appear to
be more conclusive. We have shown that SN 2002cx-like supernovae can
be explained well by pure deflagrations of Chandrasekhar-mass WDs that 
leave behind a bound WD. SN 1991bg-like events, in contrast, can be
explained by a violent merger of two WDs of almost equal
mass around 0.9\,\msun. Finally, in our simulations we did not find
an explanation of the superluminous SNe~Ia. Neither the merger of two
massive WDs nor the explosion of a rapidly-rotating
super-Chandrasekhar-mass WD can reproduce the high luminosity and
rather normal expansion velocity at the same time. Here, one might
speculate that not all the luminosity comes from radioactive decay.

\begin{acknowledgments}
We are grateful to MPA's SN Ia group  and, in particular, to 
Franco Ciaraldi-Schoolman, Michael Fink, R\"udiger Pakmor, 
Ivo Seitenzahl, Stuart Sim, and 
Stefan Taubenberger for many inspiring discussions and their 
valuable contributions to the work that is presented in this review.
This work was supported by the Deutsche Forschungsgemeinschaft via
the Transregional Collaborative Research Center TRR~33 and the
Excellence Cluster EXC~153. The research of F.K.R. is supported by 
the Emmy Noether Program (RO 3676/1-1) of the Deutsche
Forschungsgemeinschaft and by the ARCHES prize of the
German Federal Ministry of Education and Research (BMBF). The simulations 
were performed at the J\"ulich supercomputing center (grants 
PRACE026, PRACE042 and HMU014), the Computing Center Garching 
of the Max-Planck-Gesellschaft and NCI at the ANU. 
\end{acknowledgments}

\bibliographystyle{apsrev}
\bibliography{astrofritz}


\end{document}